\documentclass[aps,reprint,amsmath,amssymb,onecolumn,superscriptaddress]{revtex4-2}
\usepackage{graphicx}
\usepackage{dcolumn}
\usepackage{bm}
\usepackage{subfigure}
\usepackage{float}
\usepackage{color}
\usepackage{mathrsfs}
\usepackage{amstext}
\usepackage{booktabs}
\usepackage{siunitx}
\usepackage{lineno}
\begin{document}


\title{Scaling Enhancement in Quantum Metrology via Indefinite-Time-Direction Encoding}


\author{Binke Xia}
\affiliation{State Key Laboratory of Photonics and Communications, Institute for Quantum Sensing and Information Processing, School of Automation and Intelligent Sensing, Shanghai Jiao Tong University, Shanghai 200240, China}
\affiliation{QICI Quantum Information and Computation Initiative, School of Computing and Data Science, The University of Hong Kong, Pokfulam Road, Hong Kong, China}
\author{Jingzheng Huang}
\email{jzhuang1983@sjtu.edu.cn}
\affiliation{State Key Laboratory of Photonics and Communications, Institute for Quantum Sensing and Information Processing, School of Automation and Intelligent Sensing, Shanghai Jiao Tong University, Shanghai 200240, China}
\affiliation{Hefei National Laboratory, Hefei 230088, China}
\affiliation{Shanghai Research Center for Quantum Sciences, Shanghai 201315, China}
\author{Yuxiang Yang}
\email{yxyang@hku.hk}
\affiliation{QICI Quantum Information and Computation Initiative, School of Computing and Data Science, The University of Hong Kong, Pokfulam Road, Hong Kong, China}
\author{Guihua Zeng}
\affiliation{State Key Laboratory of Photonics and Communications, Institute for Quantum Sensing and Information Processing, School of Automation and Intelligent Sensing, Shanghai Jiao Tong University, Shanghai 200240, China}
\affiliation{Hefei National Laboratory, Hefei 230088, China}
\affiliation{Shanghai Research Center for Quantum Sciences, Shanghai 201315, China}



\begin{abstract}
The precision limit in quantum metrology, quantified by the root-mean-square error of parameter estimation, is conventionally expected to improve at most linearly with the total interrogation time $T$ and with the number $N$ of queried quantum gates. Although several metrological schemes have been shown to achieve precision scaling faster than linear in $T$ and $N$, they typically rely on unbounded probe-side information resources, usually qualified by an increasingly large variance of the parameter generator. This requirement complicates the interpretation of the resulting scaling advantage and poses substantial technical challenges. In this work, we employ an indefinite-time-direction encoding process to achieve a nonlinear-scaling enhancement of the precision limit. Rather than relying on increasingly informative probe states, our method converts controllable noncommuting encoding operations into metrological gain. Experimentally, we implement this protocol for angular-rotation measurement in a quantum optical system and demonstrate a nonlinear-scaling improvement in practical precision without using probe-side information resources. These results establish a practical framework for surpassing conventional linear-scaling precision limits in quantum metrology and provide new insights into precision enhancement in realistic quantum metrological and sensing applications.
\end{abstract}


\maketitle

\section{Introduction}
Quantum metrology plays a pivotal role in both fundamental physics and advanced technologies, including the Laser Interferometer Gravitational-Wave Observatory (LIGO) \cite{LIGO_2011,PhysRevLett.123.231107,Abe_2021}, navigation systems \cite{RevModPhys.87.637,Bongs_2019,s21165568}, and biochemical applications \cite{doi:10.1021/acs.accounts.5b00484,Aslam_2023}. A central task in quantum metrology is to determine the ultimate precision limit for estimating unknown parameters. Traditionally, the Heisenberg limit has been regarded as the fundamental scaling in quantum parameter estimation \cite{Giovannetti_2004,PhysRevLett.96.010401,Giovannetti_2011}, corresponding to a precision that improves linearly with either the interrogation time $T$ \cite{PhysRevLett.115.110401} (scaling as $1/T$) or the number $N$ of queried quantum gates/probes \cite{PhysRevLett.65.1348,Mitchell_2004,doi:10.1126/science.1188172,PhysRevA.96.012117,PhysRevLett.123.040501} (scaling as $1/N$).

Recent studies have demonstrated metrological schemes exhibiting precision scalings beyond the conventional Heisenberg form, often referred to as ``super-Heisenberg'' scalings \cite{PhysRevLett.98.090401,PhysRevLett.100.220501,Napolitano_2011,Pang_2017,PhysRevLett.126.070503,PhysRevLett.123.110501,PhysRevLett.127.060501,PhysRevLett.124.190503,Yin_2023}, in which the estimation error decreases faster than linearly with respect to $T$ or $N$. However, such schemes typically rely on unbounded probe-related resources, and are therefore subject to interpretational controversies \cite{PhysRevLett.105.180402,PhysRevX.2.041006,PhysRevX.8.021022}. For instance, the nonlinear interactions among probes in \cite{PhysRevLett.98.090401,PhysRevLett.100.220501,Napolitano_2011} induce a nonlinear growth in the universal energy cost, while the time-dependent Hamiltonian controls employed in \cite{PhysRevLett.123.110501,PhysRevLett.127.060501} lead to an unbounded dynamical variation of the probe state. In general, the corresponding informative probe states are challenging to realize and scale experimentally.

To surpass the conventional linear-scaling precision limit without increasing the probe-side information resource, we employ an indefinite-time-direction (ITD) strategy enabled by the quantum switch \cite{Chiribella_2022}, which encodes a coherent superposition of noncommuting shifts on the probe state. ITD techniques have demonstrated their ability to enhance various quantum information processing tasks, including quantum discrimination, quantum games \cite{PhysRevLett.132.160201, PhysRevResearch.6.023071}, and quantum metrology \cite{doi:10.1126/sciadv.adm8524,Agrawal2025indefinitetime}. Unlike previous ITD-based metrological schemes, our protocol does not require redesigning the unitary process that loads the parameter of interest, and achieves quadratic-scaling precision ($\propto 1/T^2$ and $\propto 1/N^{2}$) without using increasingly large probe-side information resources. Notably, although the ITD encoding operation modulates the probe state, it does not increase the probe-side information resource. Rather, the ITD strategy converts a controllable noncommuting encoding into metrological gain that is fully mapped on the quantum-switch ancilla for readout.

Experimentally, we implement the ITD encoding using the orbital angular momentum (OAM) and spin of photons \cite{Cimini_2023} to estimate angular rotations. Specifically, Q-plates \cite{Ambrosio_2013} are employed to implement the ITD OAM encoding for photons, while the photon spin state serves as the quantum-switch ancilla. Although OAM provides a manipulable high-dimensional Hilbert space, the probe states involved in our experiments themselves carry zero information resource for angular-rotation estimation. Instead, the ITD operation converts the noncommuting OAM encoding into metrological gain that is fully transferred to the photon spin state. Finally, by reading out the photon spin state, we demonstrate that the precision limit for angular-rotation measurement improves quadratically with both the interrogation time $T$ and the number $N$ of queried gates. Our experimental results confirm that the proposed scheme can achieve the desired precision enhancement efficiently, and further highlight the potential of utilizing OAM in quantum optical metrology. Consequently, our findings establish a practical framework for achieving nonlinear-scaling precision enhancement in quantum metrology and open up potential applications in realistic metrological and sensing tasks within quantum optical systems.

\section{Results}
\subsection{Standard quantum metrology}
\begin{figure}[htb]
	\centering
	\includegraphics[width=0.5\linewidth]{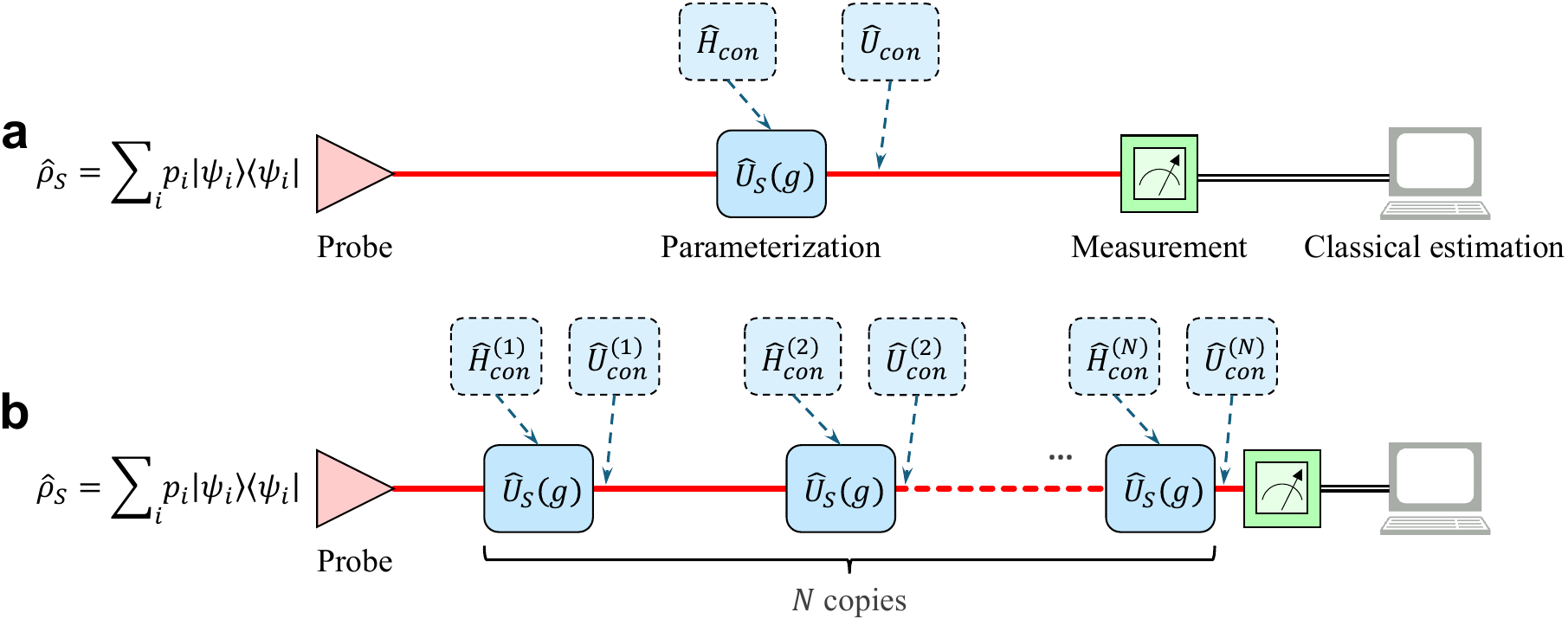}
	\caption{\label{fig:1} Schematic of standard quantum metrological schemes. \textbf{\textsf{a}} Standard quantum metrological scheme with a single-pass parameterizing process. $\hat{H}_{con}$ and $\hat{U}_{con}$ denotes the possible Hamiltonian and unitary controls applied to the probe state. \textbf{\textsf{b}} Standard quantum metrological scheme with $N$ queries of identical parameterizing processes. $\hat{H}_{con}^{(i)}$ and $\hat{U}_{con}^{(i)}$ denotes the possible Hamiltonian and unitary controls applied to the probe state during $i$-th parameterizing process.}
\end{figure}

To be clear, we first investigate the standard quantum metrological scheme, as depicted in Fig.~\ref{fig:1}\textbf{\textsf{a}}. This standard scheme consists of four stages: the preparation of the initial probe state $\hat{\rho}_{S}=\sum_{i}p_{i}|\psi_{i}\rangle\langle\psi_{i}|$, the parameterization for loading the unknown parameter $g$ into the prepared state, the practical measurement on the final state, and the estimation for determining the unknown parameter. In this work, we consider a unitary parameterizing process, wherein the unknown parameter $g$ is loaded into the quantum state $\hat{\rho}_{S}$ through the unitary evolution $\hat{U}_{S}(g)$. This evolution is generated by a parameter-dependent Hamiltonian $\hat{H}_{S}(g)$, which could in general be time dependent, acting for a duration $T_{S}$. To characterize the $g$-dependence of this dynamics, we introduce the dynamical characteristic operator $\hat{V}_{S}=\partial_{g}\hat{H}_{S}(g)$. In general, to enhance the metrological performance, we consider sequential application of $N$ identical parameterizing processes to the probe state, as depicted in Fig.~\ref{fig:1}\textbf{\textsf{b}}. The entire evolution of the probe after querying the parameterizing process $N$ times is then given by $\hat{U}_{S}^{(N)}=[\hat{U}_{S}(g)]^{N}$.

In this work, we use the quantum Fisher information (QFI) as a figure of merit \cite{Liu_2019,PhysRevApplied.13.034023, Xia:22, Xia_2023}, whose inverse determines the ultimate precision limit in estimating the unknown parameter $g$ from the parameterized state $\hat{\rho}_{S}(g)$. Subsequently, we can derive that the QFI in estimating $g$ from $\hat{\rho}_{S}^{(N)}(g)=\hat{U}_{S}^{(N)}\hat{\rho}_{S}[\hat{U}_{S}^{(N)}]^{\dagger}$ is upper bounded by (see the Supplemental Materials for details)
\begin{equation}
	\mathcal{F}_{S}^{(N)}(g) \le 4N\sum_{j=1}^{N}T_{S}\int_{0}^{T_{S}}\Delta\bar{V}_{S}^{2}(j,t)\,\mathrm{d}t, \label{eq:1}
\end{equation}
where $\Delta\bar{V}_{S}^{2}(j,t)$ denotes the average variance of $\hat{V}_{S}$ with respect to the probe state at time $t$ in the $j$-th parameterizing process. Explicitly, $\Delta\bar{V}_{S}^{2}(j,t)=\sum_{i}p_{i}[\langle\psi_{i}(j,t)|\hat{V}_{S}^{2}|\psi_{i}(j,t)\rangle-\langle\psi_{i}(j,t)|\hat{V}_{S}|\psi_{i}(j,t)\rangle^{2}]$, where $|\psi_{i}(j,t)\rangle=\hat{U}_{S}(0\to t)[\hat{U}_{S}(g)]^{j-1}|\psi_{i}\rangle$ is the evolved eigenstate of the initial probe. Eq.~(\ref{eq:1}) shows that the maximum QFI in estimating parameter $g$ is completely determined by the dynamical variation of the probe state with respect to $\hat{V}_{S}$ during the metrological process, which is characterized by the average operator fluctuation $\Delta\bar{V}_{S}(j,t)=\sqrt{\Delta\bar{V}_{S}^{2}(j,t)}$. The maximum attainable value of $\Delta\bar{V}_{S}(j,t)$ thus quantifies the largest probe-related (dynamical) resource available in the entire parameterizing process. We impose a constraint on this maximum dynamical resource as
\begin{equation}
	\mathcal{V} = \max_{\hat{\rho}_{S}(j,t)} \Delta\bar{V}_{S} \le V, \label{eq:2}
\end{equation}
where $\hat{\rho}_{S}(j,t)$ is the intermediate probe state at time $t$ in the $j$-th parameterizing process, $\Delta\bar{V}_{S}$ is the corresponding average uncertainty of $\hat{V}_{S}$ and $V$ is a fixed constant. Hamiltonian controls $\hat{H}_{con}$ and unitary controls $\hat{U}_{con}$ that do not violate this resource constraint are allowed throughout the entire metrological process. Under the constraint in Eq.~(\ref{eq:2}), the QFI in eq. (\ref{eq:1}) is further bounded as $\mathcal{F}_{S}^{(N)}(g)\le 4V^{2}N^{2}T_{S}^{2}$, which leads to an ultimate precision limit
\begin{equation}
	{\delta g}_{S}^{(N)} \ge \frac{1}{2NTV\sqrt{M}}, \label{eq:3}
\end{equation}
where $T=T_{S}$ is the time length of each single parameterizing process and $M$ is the number of independent measurements. In particular, for a non-informative probe, defined as a probe whose maximum dynamical resource $\mathcal{V}=0$, the parameter $g$ becomes inestimable within the standard scheme. Notably, the above bound applies to both finite- and infinite-dimensional systems.

Consequently, when the probe’s dynamical resource is bounded, the ultimate precision in standard quantum metrological schemes scales at most linearly with the total evolution time $T$ or with the number $N$ of queried quantum gates: $\delta g_{S}\propto 1/T$ for a single query ($N=1$) and $\delta g_{S}\propto 1/N$ for fixed total time ($T=1$). By contrast, ``super-Heisenberg" schemes typically rely on unbounded dynamical resources. For example, schemes based on nonlinear interactions \cite{PhysRevLett.98.090401,PhysRevLett.100.220501,Napolitano_2011} require an $N$-dependent dynamical resource $\mathcal{V}\propto N$, while time-dependent Hamiltonian controls in \cite{PhysRevLett.123.110501,PhysRevLett.127.060501} employ a $T$-dependent dynamical resource $\mathcal{V}\propto T$. Although indefinite causal order (ICO) does not, in principle, require unbounded resources \cite{PhysRevLett.124.190503}, its experimental implementation \cite{Yin_2023} still relies on informative probe states with $\mathcal{V}>0$. (See the Methods part for details.)

\subsection{ITD-encoding quantum metrology}
\begin{figure}[htb]
	\centering
	\includegraphics[width=0.5\linewidth]{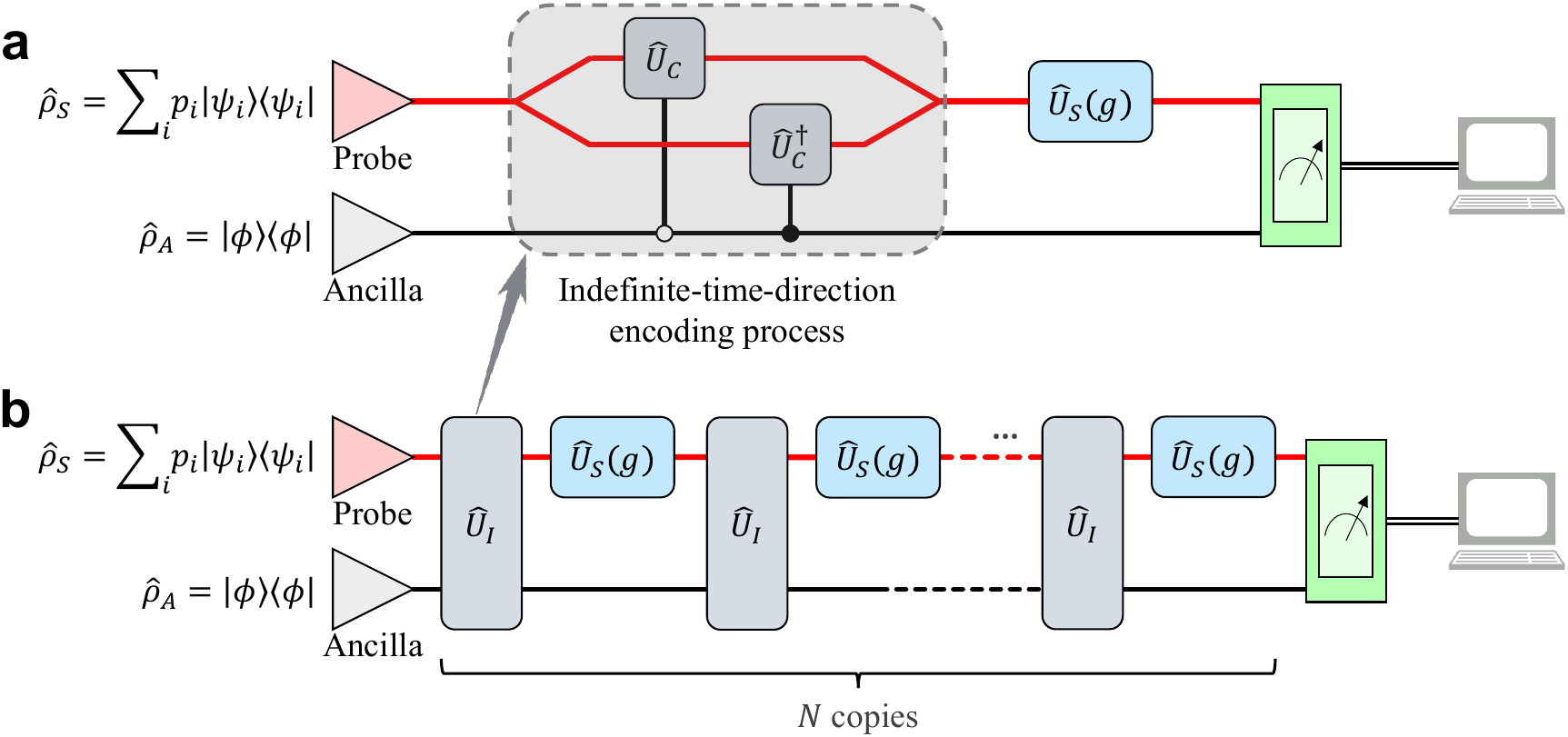}
	\caption{\label{fig:2} Schematic of ITD-encoding metrological schemes. \textbf{\textsf{a}} Quantum metrological scheme with an ITD encoding process. $\hat{U}_{C}$ denotes the conjugated encoding process with definite time direction. The ancilla is a qubit to implement the quantum switch. \textbf{\textsf{b}} Quantum metrological scheme with $N$ queries of identical ITD encoding processes and parameterizing processes.}
\end{figure}

To surpass the linear-scaling limits without relying on informative probe states, we devise an encoding process with an indefinite time direction (ITD) prior to the parameterizing process. The ITD encoding process is implemented via a two-level ancilla $\hat{\rho}_{A}=|\phi\rangle\langle\phi|$ with eigenstates $|0\rangle$ and $|1\rangle$, which serves as the quantum switch, as depicted in Fig.~\ref{fig:2}\textbf{\textsf{a}}. The corresponding evolution is described by $\hat{U}_{I}=\hat{U}_{C}\otimes|0\rangle\langle 0|+\hat{U}_{C}^{\dagger}\otimes|1\rangle\langle 1|$, where $\hat{U}_{C}=\exp(-\mathrm{i}\hat{H}_{C}T_{C})$ denotes the encoding process with a definite time direction and $T_{C}$ is its duration. In this way, the probe state $\hat{\rho}_{S}$ undergoes a coherent superposition of forward- and backward-time evolutions.

In our scheme, we impose that the encoding Hamiltonian satisfies $[\hat{H}_{C},\hat{V}_{S}]=\pm\mathrm{i}$ and the commutator $[\hat{H}_{C},\hat{H}_{S}]$ commutes with both $\hat{H}_{C}$ and $\hat{H}_{S}$ at any time. As a result, this process encodes shifts of the operator $\hat{V}_{S}$ on the probe state but does not increase the dynamical resource associated with the probe, i.e. the average uncertainty of $\hat{V}_{S}$. Nevertheless, we can calculate that the QFI in estimating $g$ from the final joint state of probe and ancilla is given by (see the Supplemental Materials for calculation details)
\begin{equation}
	\mathcal{F}_{I}^{(N)}(g) = \mathcal{F}_{S}^{(N)}(g)+T_{C}^{2}T_{S}^{2}\left(N^{2}+N\right)^{2}\left(1-\bar{\sigma}_{z}^{2}\right), \label{eq:4}
\end{equation}
when the identical ITD encoding and parameterizing processes are queried $N$ times. Here, $\bar{\sigma}_{z}=\langle\phi|\hat{\sigma}_{z}|\phi\rangle$, with $\hat{\sigma}_{z}=|0\rangle\langle 0|-|1\rangle\langle 1|$ denoting the Pauli-$Z$ operator acting on the ancilla. The total evolution time of a single query for the ITD encoding process and the parameterizing process is $T=T_{C}+T_{S}$. By allocating this time equally between the two processes, $T_{C}=T_{S}=T/2$, the an additional increment in $\mathcal{F}_{I}^{(N)}(g)$ scales as $T^{4}$ over the QFI $\mathcal{F}_{S}^{(N)}(g)$ of the standard scheme. In addition, by initializing the ancilla state as $|\phi\rangle=(|0\rangle+|1\rangle)/\sqrt{2}$, the ultimate precision limit in our scheme is given by
\begin{equation}
	{\delta g}_{I}^{(N)} \ge \frac{4}{\sqrt{M\left[16V^{2}T^{2}N^{2}+T^{4}(N^{2}+N)^{2}\right]}}, \label{eq:5}
\end{equation}
when the maximum dynamical resource on the probe state is bounded by $V$. This result indicate that, even when non-informative probes with $\mathcal{V}=0$ are employed, our scheme can still achieve a precision limit with nonlinear scaling, $\delta g_{I}^{(N)} \propto 1/T^{2}(N^{2}+N)$, with respect to the evolution time $T$ and the number $N$ of queried metrological processes. Therefore, the scaling enhancement achieved by our scheme does not rely on intrinsic information resources carried by the probes themselves. Rather, it originates from the quantum resources introduced by the ITD encoding, namely quantum superposition and noncommutativity. Specifically, the ITD strategy directly converts a controllable noncommuting encoding operation into a metrological information gain for estimating the parameter of interest. This gain is independent of the probe state and is fully transferred to, and read out from, the quantum-switch ancilla.

In principle, our encoding process does displace the probe state with respect to the operator $\hat{V}_{S}$. Although this operation does not modify its average uncertainty (i.e., the dynamical resource), its physical implementation may entail additional energy costs. Nevertheless, in quantum optical systems, several photonic degrees of freedom, such as spatial modes and orbital angular momentum (OAM), offer manipulable high-dimensional Hilbert spaces without increasing the photon energy. By leveraging these properties, the nonlinear-scaling enhancements predicted by our scheme can be implemented without incurring additional energy costs.

\subsection{Experimental implementation for angular-rotation measurement}
For experimental demonstration, we investigate a practical metrological scenario of angular rotation in the quantum optical system. The unitary evolution of the parameterizing process is given as $\hat{U}_{S}=\exp(-\mathrm{i}gT_{S}\hat{L}_{z})$, where $\hat{L}_{z}$ is the OAM operator of photons and $gT_{S}=\alpha$ is the rotation angle. In our experimental settings, we modulate different values of the rotation angle $\alpha$ on photons to achieve the desired values of the unknown parameter $g$ or the evolution time length $T_{S}$, assuming one of their values is fixed. Specifically, the evolution time length $T_{S}$ will be directly determined by the rotation angle $\alpha$ when the value of the unknown parameter $g$ is fixed as a constant, and vice versa. Therefore, the encoding process ought to shift the OAM of photons, which can be designed as $\hat{U}_{C}=\exp(-\mathrm{i}T_{C}\hat{\varphi})$, wherein the evolution time length $T_{C}=m$ is the topological charge in practice and the angular position operator $\hat{\varphi}$ satisfies $[\hat{\varphi},\hat{L}_{z}]=\mathrm{i}$. This unitary evolution of the encoding process introduces an extra spiral phase on the photons, thereby shifting the OAM of the photons by a value of $m$.

\begin{figure}[htb]
	\centering
	\includegraphics[width=0.5\linewidth]{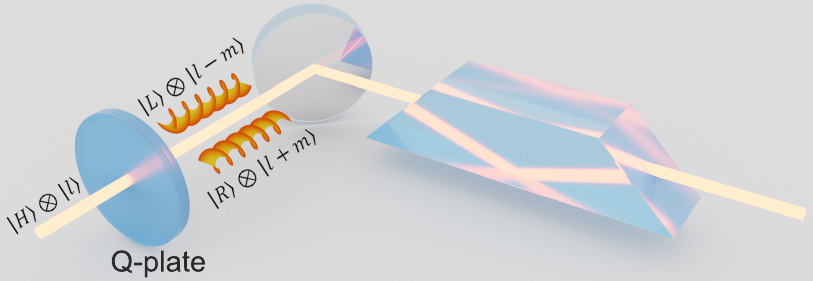}
	\caption{\label{fig:3} Experimental implementation of the ITD encoding process and the parameterizing process. The Q-plate encodes a superposition of opposite OAM shifts. The Dove prism loads the angular rotation parameter.}
\end{figure}

\begin{figure*}[htb]
	\centering
	\includegraphics[width=0.9\linewidth]{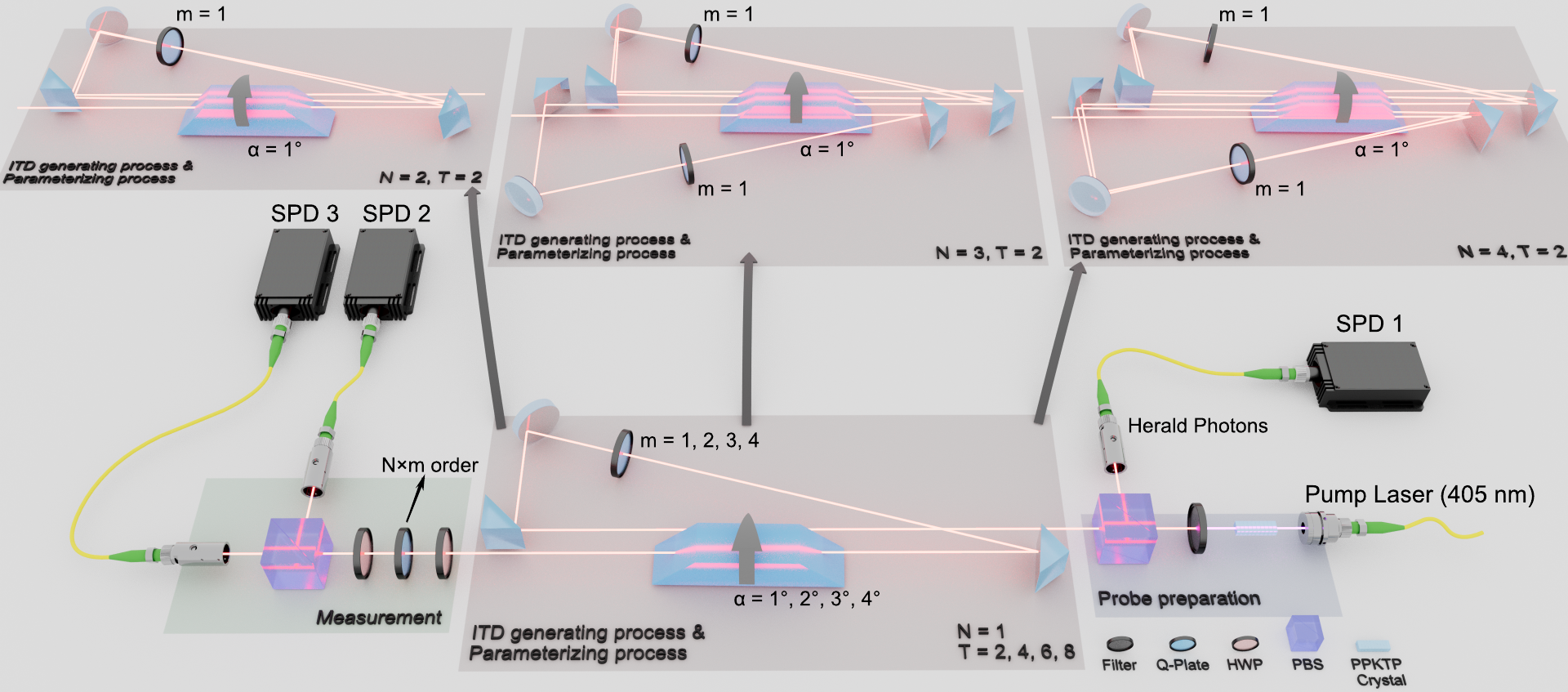}
	\caption{\label{fig:4} Experimental setup. The SPDC process in the PPKTP crystal converts a $\SI{405}{\nm}$ photon into an $\SI{810}{\nm}$ single-photon pair with orthogonal polarization. Idler photons are separated by a PBS to herald signal photons. The signal photon serves as the probe, and its polarization state serves as the ancilla. Different Q-plate orders and Dove prism rotation angles implement various evolution times $T$. By adjusting right-angle prism mirrors, photons pass through the Q-plate and Dove prism multiple times, achieving various iteration numbers $N$. Projective measurement is conducted using a PBS, two HWPs, and an $N\times m$-order Q-plate. Photon numbers under orthogonal projections are detected via coincidence counting between SPD 2 and SPD 1, and between SPD 3 and SPD 1.}
\end{figure*}

In practice, we use the OAM state of photons as the probe, initially prepared in a Gaussian-profile beam with zero OAM. Since this probe is an OAM eigenstate, it carries no information resource (i.e., $\mathcal{V} = \max\Delta\bar{L}_{z} = 0$), and this condition is maintained throughout the experiment. The spin state of photons serves as the ancilla, with the right-handed circular polarization state $|R\rangle$ and left-handed circular polarization state $|L\rangle$ forming an orthogonal basis. The ancilla state is initialized as horizontal polarization state $|H\rangle=\frac{1}{\sqrt{2}}(|R\rangle+|L\rangle)$. As depicted in Fig.~\ref{fig:3}, we use a Q-plate \cite{Ambrosio_2013} and a Dove prism to implement the ITD encoding process the parameterizing process in experiment. The Q-plate introduces a pair of opposite OAM shifts on the probe state based on the spin state of photons. Specifically, the $m$-order Q-plate flips the spin state of ancilla, meanwhile shifts the OAM of the probe state by a value of $-m$ with right-handed spin $|R\rangle$ and a value of $m$ with left-handed spin $|L\rangle$. To compensate this spin flip, we also insert a mirror between the Q-plate and the Dove prism. 

Fig.~\ref{fig:4} delineates the configuration of our experimental apparatus. To initialize the probe and ancilla, we generate single-photon pairs through a degenerated type-II spontaneous parametric down-conversion (SPDC) process within a periodically poled potassium titanyl phosphate (PPKTP) crystal, pumped by a $\SI{405}{\nm}$ continuous wave (cw) laser. Due to the SPDC process inside the PPKTP crystal, the single-photon pairs work at $\SI{810}{\nm}$, then the signal photon and idler photon are separated by a polarizing beam splitter (PBS). Thus the polarization of the signal photon is initialized as $|H\rangle$. The idler photon is finally measured by a single photon detector (SPD) and serves as a trigger for the signal photon evolved through the experimental setup.

Fist, we fix the value of unknown parameter as $g=\pi/180$, thus various rotation angles $\alpha$ represents the distinct parameterizing evolution times $T_{S}$. Since rotation angle $\alpha=gT_{S}$, a rotation angle of $\alpha=\ang{1}$ corresponds to a normalized evolution time of $T_{S}=1$ in this case. Additionally, the evolution time $T_{C}$ of the encoding process is represented by the order $m$ of the Q-plate ($T_{C}=m$), with $m=1$ corresponding to a normalized evolution time $T_{C}=1$. Here, we designate the dimensionless evolution time $T_{S}=T_{C}=T/2$. To estimate the unknown parameter from the final state, a practical projective measurement is devised for the experiment. As depicted in Fig.~\ref{fig:4}, a half-wave plate (HWP) followed by a $m$-order Q-plate are used to eliminate the topological charges on probe state. Subsequent to this, another HWP, with its optical axis set at an angle of $\ang{22.5}$ to the horizontal plane, followed by a PBS, are employed to project the final state of ancilla onto the $\ang{45}$ linearly polarized state $|+\rangle$ and the $-\ang{45}$ linearly polarized state $|-\rangle$. We denote the projective probabilities of these two orthogonal projections as $P_{+}$ and $P_{-}$, respectively. Using the classical parameter estimation theory, the classical Fisher information (CFI) in estimating parameter $g$ can be calculated as $F(g)=T^{4}/4$, where $T=T_{C}+T_{S}$ is the total time length of a single metrological process. (See the Methods part for calculation details of CFI.) Therefore, the experimental precision in estimating parameter $g$ is given by
\begin{equation}
	\delta g_{\mathrm{exp}} \ge \frac{2}{T^{2}\sqrt{M}}, \label{eq:6}
\end{equation}
where $M$ is the number of independent measurements (the number of measured photons).

\begin{figure}[htb]
	\centering
	\includegraphics[width=0.45\linewidth]{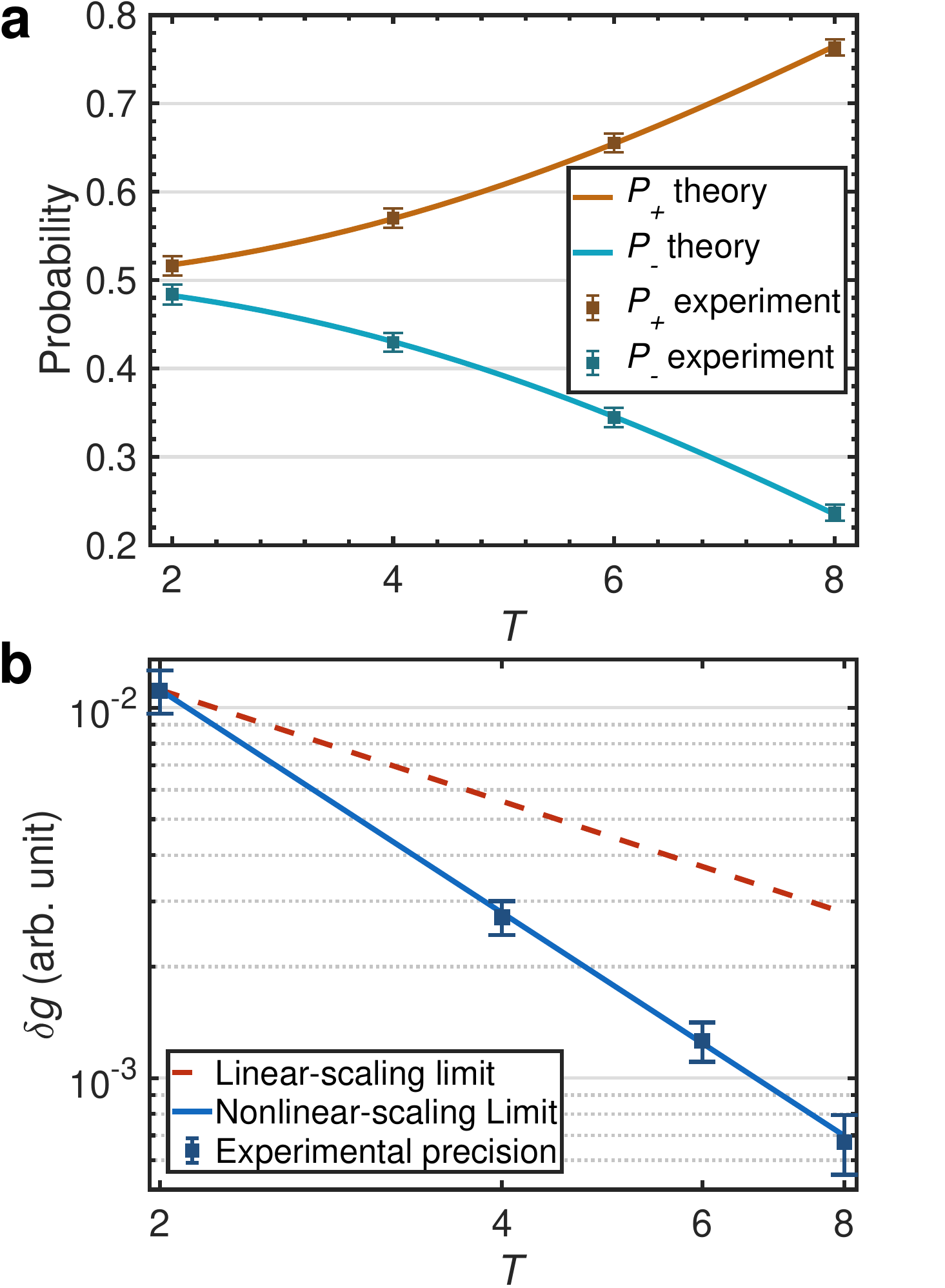}
	\caption{\label{fig:5} Experimental results with different evolution time $T=T_{C}+T_{S}$ for a single query ($N=1$). \textbf{\textsf{a}} Projective probabilities $P_{+}$ and $P_{-}$. The brown and cyan solid lines denote the corresponding theoretical predictions, and the brown and cyan squares with error bars denote the experimental data. \textbf{\textsf{b}} RMSEs in estimating $g$ with different time $T$. The blue solid line represents the nonlinear-scaling precision limit given by Eq.~(\ref{eq:6}). The red dashed lines gives a reference linear-scaling precision limit for the standard quantum metrological scheme with $V=1/2$. The blue squares with error bars give the experimental RMSEs for estimating the corresponding unknown parameters.}
\end{figure}

In the experiment, we perform the coincidence counting between SPD 1 and SPD 2, SPD 3 to get the photon number under the two orthogonal projections. Subsequently, we estimate parameter $g$ from the detected photon numbers under these two projections and calculate the corresponding root-mean-square error (RMSE) from these estimates. As shown in Fig.3, we first conduct the experiments of a single metrological process with a total time $T=T_{C}+T_{S}$ of 2, 4, 6, and 8. Specifically, the different time lengths are achieved by rotating the photon profile with an angle $\alpha=\ang{1}$, $\ang{2}$, $\ang{3}$, and $\ang{4}$ meanwhile choosing the topological order $m$ of the Q-plate as 1, 2, 3, and 4, separately. Fig.~\ref{fig:5}\textbf{\textsf{a}} displays the experimental results of projection probabilities $P_{+}$ and $P_{-}$ with their respective error bars, and plots the theoretical results for comparison. Subsequently, we calculate the RMSE in estimating the parameter $g$, where the results are illustrated in Fig.~\ref{fig:5}\textbf{\textsf{b}} with error bars. For comparison, the theoretical precision limit of Eq.~(\ref{eq:6}) is plotted using a solid line, and the linear-scaling precision limit (${\delta g}_{S}=1/T\sqrt{M}$) of the standard quantum metrological scheme with maximum dynamical resource $V=1/2$ is plotted using a dashed line.

\begin{figure}[htb]
	\centering
	\includegraphics[width=0.45\linewidth]{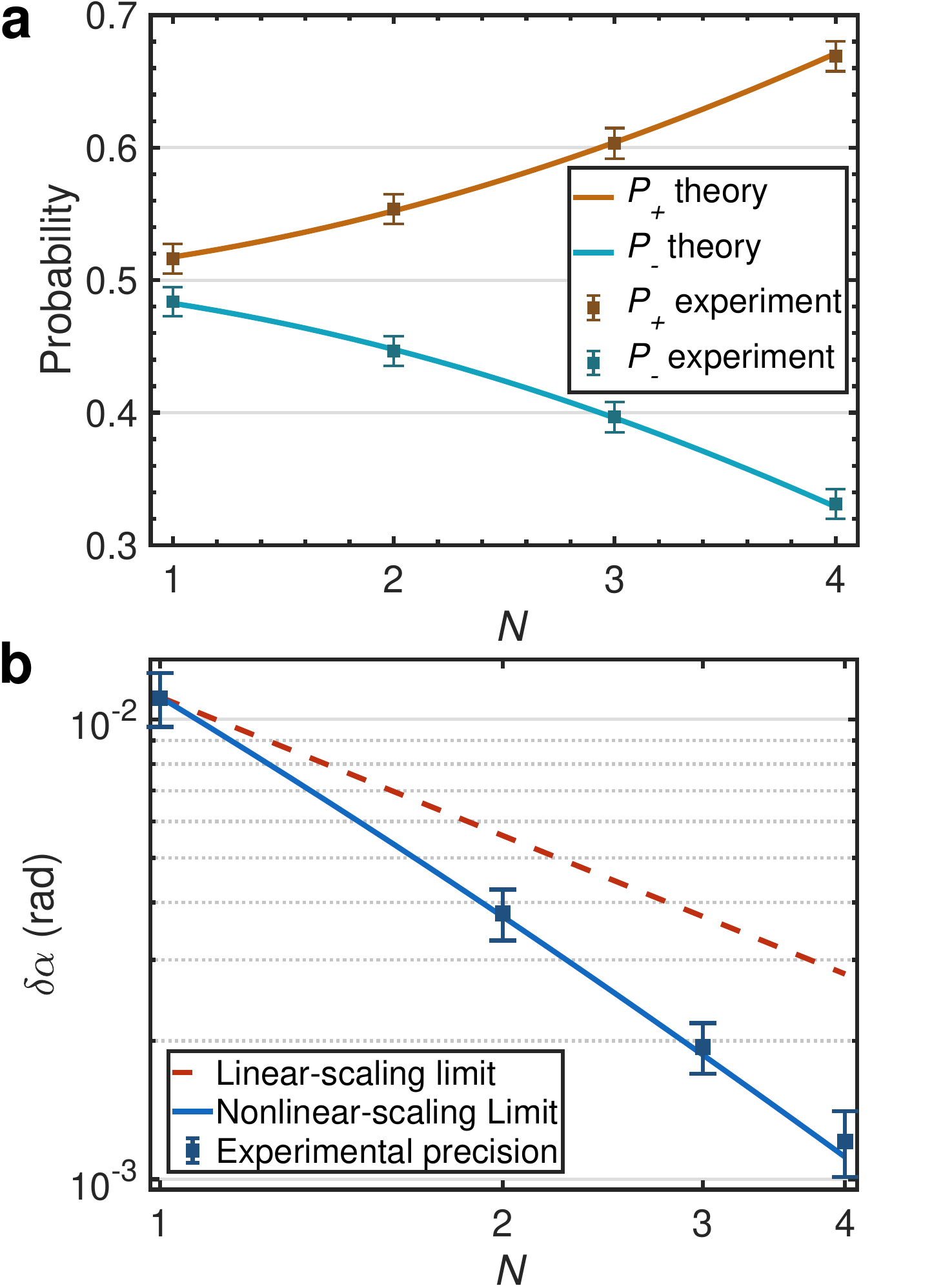}
	\caption{\label{fig:6} Experimental results with different number $N$ of queries metrological process for a fixed time ($T_{C}=T_{S}=1$). \textbf{\textsf{a}} Projective probabilities $P_{+}$ and $P_{-}$. The brown and cyan solid lines denote the corresponding theoretical predictions, and the brown and cyan squares with error bars denote the experimental data. \textbf{\textsf{b}} RMSEs in estimating $\alpha$ with different number $N$. The blue solid line represents the nonlinear-scaling precision limit given by Eq.~(\ref{eq:7}). The red dashed line gives a reference linear-scaling precision limit for the standard quantum metrological scheme with $V=1$. The blue squares with error bars give the experimental RMSEs for estimating the corresponding unknown parameters.}
\end{figure}

Subsequently, we sequentially apply this single metrological process comprising the ITD encoding and the parameterizing $N$ times to each probe (single photon). We here fix the value of the dimensionless evolution time as $T_{S}=T_{C}=1$ and $T=T_{C}+T_{S}=2$, which means the topological order of the Q-plate is set to $m=1$ within our experimental settings. In this context, the unknown parameter $g$ is directly determined by the rotation angle $\alpha$ of the photon beam profile, i.e., $\alpha=gT_{S}=g$. We use experimental settings similar to those we've mentioned above to perform projective measurements on this final state, where an $N$-order Q-plate is employed in this case. Finally, we can calculate the corresponding CFI as $F^{(N)}(g)=F(\alpha)=(N^{2}+N)^{2}$, which leads to the experimental precision
\begin{equation}
	\delta g_{\mathrm{exp}} ^{(N)} = \delta\alpha \ge \frac{1}{(N^{2}+N)\sqrt{M}}. \label{eq:7}
\end{equation}

We conduct the experiments of an iteration number $N$ of 1, 2, 3, and 4 by moving the right-angle prism mirrors in our experimental system, the corresponding experimental light paths are shown in the upper half part of Fig.~\ref{fig:4}. We illustrate the experimental results of $P_{+}$ and $P_{-}$ with their respective error bars in Fig.~\ref{fig:6}\textbf{\textsf{a}}, and plot their theoretical predictions for comparison. The results of the RMSE in estimating parameter $\alpha$ are illustrated in Fig.~\ref{fig:6}\textbf{\textsf{b}}, with error bars. For comparison, the theoretical precision limit of Eq.~(\ref{eq:7}) is plotted using a solid line, and the linear-scaling precision limit ($\delta\alpha=1/(2N\sqrt{M})$) of the standard quantum metrological scheme with maximum dynamical resource $V=1$ is plotted using a dashed line.

\section{Discussion and Conclusion}
In our experimental implementation, we rely on the noncommutative relation between angle operator $\hat{\varphi}$ and OAM operator $\hat{L}_{z}$ to realize the noncommuting ITD encoding operation. However, the commutation relation $[\hat{\varphi},\hat{L}_{z}]=\mathrm{i}$ holds only when the domain of $\hat{\varphi}$ is restricted to any open interval within $[0, 2\pi]$. This is because $\varphi$ is not differentiable at the boundaries of $[0, 2\pi]$, so that $\hat{L}_{z}=-\mathrm{i}\partial_{\varphi}$ is not applicable at these points. As discussed in \cite{PhysRevA.73.052104}, this problem can be avoided by considering the functions $\cos\hat{\varphi}$ and $\sin\hat{\varphi}$ as basic observables, instead of $\hat{\varphi}$ itself, since the latter is not a good observable either classically or quantum theoretically. In our experimental scheme, we actually consider the commutation relation between the OAM operator $\hat{L}_{z}$ and the unitary operator $\hat{U}_{C}$, wherein $\mathrm{e}^{\mathrm{i}\hat{\varphi}}=\cos\hat{\varphi}+\mathrm{i}\sin\hat{\varphi}$ is taken as the basic observable. Thus, the complications associated with the discontinuity of the angle operator $\hat{\varphi}$ do not arise within our framework.

In summary, by devising a ITD-encoding protocol enabled by the quantum switch, we have established a general framework for achieving nonlinear-scaling precision limits in quantum metrology that does not rely on information resources intrinsic to the probes themselves. We have demonstrated this scaling enhancement both theoretically and experimentally using probes with zero intrinsic information resource. Experimentally, we implement ITD encoding via the OAM of photons to estimate angular rotations in a quantum optical setup. The OAM operations access high-dimensional Hilbert spaces while conserving photon energy. Consequently, these results not only advance the ultimate precision limits attainable in quantum metrology, but also hold significant promise for practical quantum optical information processing, including applications in optical sensing \cite{doi:10.1126/sciadv.adm8524} and communication \cite{Willner:15} enabled by OAM encoding.

\section{Methods}
\subsection{Experimental materials}
We used a PPKTP crystal from Raicol Crystals with dimensions $\qtyproduct[product-units=power]{1 x 2 x 5}{\mm}$. The Q-plates employed in our experiments were custom-made by LBTEK. To ensure that photons could traverse the Dove prism during each sequential evolution, we selected a Dove prism from Thorlabs with a transverse size of $\qtyproduct[product-units=power]{30 x 30}{\mm}$. The Dove prism was mounted on a motorized goniometer equipped with a stepper motor to facilitate accurate rotation angles. Single-photon detection was carried out using Avalanche Photodiode (APD) detectors from LBTEK, operating in Geiger mode to amplify the signal of individual photons and output pulse signals to the event counter. The detected pulse signals from the three SPDs were input into the Moku:Pro from Liquid Instruments, which functioned as the event counter. An algorithm was developed to calculate the coincidence counts between SPD 2 and SPD 1, and between SPD 3 and SPD 1 from the logged timestamps provided by Moku:Pro.

To record the experimental results, we establish the detection time span for each SPD at $\SI{50}{\ms}$, and standardize the total detected photon count $\tilde{m}$ to approximately 2000, achieved by regulating the pump laser's power. This experiment was replicated 600 times, accomplished by logging the counting events of each SPD continuously for $\SI{30}{\s}$ and subsequently segregating them into 600 groups with time spans of $\SI{50}{\ms}$. This enables the calculation of projective probabilities $P_{+}$ and $P_{-}$, as well as the estimated value $\tilde{g}$ of the unknown parameter $g$ for each experimental result group. To evaluate the RMSE in estimating parameter $g$, we then divide the 600 sets of experimental results into 20 groups, each containing 30 sets of results, and calculate the RMSE of the unknown parameter $g$ for each group. The experimental results of $P_{+}$, $P_{-}$, and the RMSEs are illustrated in Fig.~\ref{fig:5} and Fig.~\ref{fig:6}.

\subsection{Dynamical resources in ``super-Heisenberg'' schemes}
Here, we provide a detailed analysis of the dynamical resources involved in different quantum metrological schemes that exhibit so-called ``super-Heisenberg'' scaling.

For the nonlinear-interaction-based schemes considered in \cite{PhysRevLett.98.090401,PhysRevLett.100.220501,Napolitano_2011}, the experimental implementation typically involves a light-matter quantum interface, with Hamiltonian $\hat{H}_{S}=\chi\hat{S}_{z}\hat{S}_{0}=\chi\hat{S}_{z}N$, where $\xi$ is the unknown parameter to be estimated. Here, $\hat{S}_{z}=\sum_{j=1}^{N}\hat{\sigma}_{z}^{(j)}$ is a collective operator describing the net polarization of the photons, and $\hat{S}_{0}$ denotes the sum of $N$ identity operators. In this case, the characteristic operator is given by $\hat{V}_{S}=\partial_{\chi}\hat{H}_{S}=\hat{S}_{z}N$. Accordingly, for $N$ photons, the total maximum dynamical resource is $\mathcal{V}^{(N)}=\max N\Delta\bar{S}_{z}=N^{2}$, which corresponds to a per-photon dynamical resource $\mathcal{V}=\mathcal{V}^{(N)}/N=N$.

For the time-dependent schemes in \cite{PhysRevLett.123.110501,PhysRevLett.127.060501}, the sensing dynamics are governed by the Hamiltonian $\hat{H}_{S}=-B[\hat{\sigma}_{x}\cos(\omega t)+\hat{\sigma}_{z}\sin(\omega t)]$, where $\omega$ is the unknown frequency to be estimated. The corresponding characteristic operator is $\hat{V}_{S}=\partial_{\omega}\hat{H}_{S}=Bt[\hat{\sigma}_{x}\sin(\omega t)+\hat{\sigma}_{z}\cos(\omega t)]=Bt\hat{\sigma}(t)$. Under optimal Hamiltonian control, this yields a maximum dynamical resource $\mathcal{V}=\max Bt\Delta\bar{\sigma}^{2}(t)=BT_{S}$.

For the ICO-based schemes in \cite{PhysRevLett.124.190503,Yin_2023}, a coherent-state probe $|\alpha\rangle$ is used to query $N$ $\hat{P}$-displacement gates $\hat{U}_{x_{j}}=\exp(-\mathrm{i}x_{j}\hat{P})$ for $j=1,2,\dots,N$ and $N$ $\hat{X}$-displacement gates $\hat{U}_{p_{j}}=\exp(-\mathrm{i}p_{j}\hat{X})$ for $j=1,2,\dots,N$ in a superposition of causal orders. The corresponding joint probe-ancilla evolution is
\begin{align}
	\hat{U}_{\mathrm{jt}} &= \exp(-\mathrm{i}N\bar{p}\hat{X})\exp(-\mathrm{i}N\bar{x}\hat{P})\otimes|0\rangle\langle 0| \nonumber \\
	&\qquad+\exp(-\mathrm{i}N\bar{x}\hat{P})\exp(-\mathrm{i}N\bar{p}\hat{X})\otimes|1\rangle\langle 1|, \label{eq:8}
\end{align}
where $\bar{x}=\sum_{j=1}^{N}x_{j}/N$ and $\bar{p}=\sum_{j=1}^{N}p_{j}/N$. The parameter to be estimated is the average geometric phase $\bar{\mathcal{A}}=\bar{x}\bar{p}$ in the ICO-based scheme. By using the Baker-Campbell-Hausdorff formula together with $[\hat{X},\hat{P}]=\mathrm{i}$, the joint evolution can be rewritten as
\begin{align}
	\hat{U}_{\mathrm{jt}} &= \exp\left(-\mathrm{i}N\bar{p}\hat{X}\otimes\hat{\mathbb{I}}-\mathrm{i}N\bar{x}\hat{P}\otimes\hat{\mathbb{I}}-\frac{N^{2}\bar{\mathcal{A}}}{2}[\hat{X},\hat{P}]\otimes\hat{\sigma}_{z}\right) \nonumber \\
	&= \exp\left(-\mathrm{i}N\bar{p}\hat{X}\otimes\hat{\mathbb{I}}-\mathrm{i}N\bar{x}\hat{P}\otimes\hat{\mathbb{I}}-\mathrm{i}\frac{N^{2}\bar{\mathcal{A}}}{2}\otimes\hat{\sigma}_{z}\right), \label{eq:9}
\end{align}
which corresponds to an equivalent joint probe-ancilla Hamiltonian
\begin{align}
	\hat{H}_{\mathrm{jt}} &= N\bar{p}\hat{X}\otimes\hat{\mathbb{I}}+N\bar{x}\hat{P}\otimes\hat{\mathbb{I}}+\frac{1}{2}N^{2}\bar{\mathcal{A}}\otimes\hat{\sigma}_{z} \nonumber\\
	&= \hat{H}_{S}+\hat{H}_{A}, \label{eq:10}
\end{align}
where $\hat{H}_{S}=N\bar{p}\hat{X}+N\bar{x}\hat{P}$ acts only on the probe and $\hat{H}_{A}=N^{2}\bar{\mathcal{A}}\hat{\sigma}_{z}/2$ acts only on the ancilla. Accordingly, the characteristic operator associated with the probe is $\hat{V}_{S}=\partial_{\bar{\mathcal{A}}}\hat{H}_{S}=N\hat{X}/\bar{x}+N\hat{P}/\bar{p}$. For a coherent-state probe, this leads to a nontrivial maximal dynamical resource per query of the displacement gate, $\mathcal{V}=\max\Delta\bar{V}_{S}/2N=[\Delta X^{2}/4\bar{x}^{2}+\Delta P^{2}/4\bar{p}^{2}+\mathrm{Cov}(\hat{X},\hat{P})/2\bar{\mathcal{A}}]^{1/2}>0$. Although this probe-related information resource does not contribute to the scaling enhancement in the ICO-based scheme, it is nevertheless required and remains nontrivial for its experimental realization.

In contrast, our scheme employs the OAM eigenstate $|0\rangle$ as the initial probe, which carries no prior information about the angular-rotation metrological process. This is because the characteristic operator is simply the OAM operator $\hat{V}_{S}=\hat{L}_{z}$, and hence $\Delta\bar{V}_{S}=\Delta\bar{L}_{z}=\sqrt{\langle 0|\hat{L}_{z}^{2}|0\rangle-\langle 0|\hat{L}_{z}|0\rangle^{2}}=0$. Although the ITD encoding process coherently encodes a pair of opposite OAM shifts onto the probe, thereby producing the joint probe-ancilla state $|-l\rangle|R\rangle+|l\rangle|L\rangle$, the reduced state of the probe alone is a maximally mixed state $(|l\rangle\langle l|+|-l\rangle\langle-l|)/2$. Therefore, even after ITD encoding, the probe itself still carries no dynamical information resource, since $\mathcal{V}=\max\Delta\bar{V}_{S}=\sqrt{(\langle l|\Delta\hat{L}_{z}^{2}|l\rangle+\langle -l|\Delta\hat{L}_{z}^{2}|-l\rangle)/2}=0$.

\subsection{CFI in experimental scheme}
The CFI for the unknown parameter $g$ with respect to the measurement probabilities $\{P_{i}|\sum_{i}P_{i}=1\}$ can be calculated by
\begin{equation}
	F(g) = \sum_{i}\frac{1}{P_{i}}\left(\frac{\partial P_{i}}{\partial g}\right)^{2}. \label{eq:11}
\end{equation}
According to the classical Cramér–Rao bound (CCRB) theory, the practical precision of estimating $g$ from the classical measurement data satisfies
\begin{equation}
	\delta g \ge \frac{1}{\sqrt{MF(g)}}, \label{eq:12}
\end{equation}
where $M$ denotes the number of independent measurements. In particular, $M$ usually denotes the number of measured photons in quantum optical experiments.

In our scheme, we employ OAM state $|\psi(l)\rangle$ of the photon as the probe state, where $\psi(l)$ denotes its wave function in the OAM representation. In the experiments, the probe state is initialized as a photon beam with a Gaussian-profile, corresponding to an OAM value of 0. The spin state of photons is employed as the ancilla, which is is initialized as $|H\rangle=\frac{1}{\sqrt{2}}(|R\rangle+|L\rangle)$ in experiment, i.e., horizontal polarization state. We use a Q-plate to realize the ITD-encoding, , which introduces a pair of opposite OAM shifts on the probe state based on the spin state of photons. Specifically, the $m$-order Q-plate flips the spin state of ancilla, meanwhile shifts the OAM of the probe state by a value of $-m$ with right-handed spin $|R\rangle$ and a value of $m$ with left-handed spin $|L\rangle$. This process can be denoted as $\mathrm{e}^{-\mathrm{i}m\hat{\varphi}}\otimes|L\rangle\langle R|+\mathrm{e}^{\mathrm{i}m\hat{\varphi}}\otimes|R\rangle\langle L|$ mathematically. To compensate the undesirable spin flip induced by Q-plate, we insert a mirror between the Q-plate and Dove prism. Experimentally, the reflection on mirror flips the spin state of photons and inverses the topological charge of probe state simultaneously, and the total internal reflection in Dove prism only inverses the topological charge of probe state. Therefore, a single query of the ITD encoding process and the parameterizing process can be denoted as $\mathrm{e}^{-\mathrm{i}\alpha\hat{L}_{z}}\mathrm{e}^{-\mathrm{i}m\hat{\varphi}}\otimes|R\rangle\langle R|+\mathrm{e}^{-\mathrm{i}\alpha\hat{L}_{z}}\mathrm{e}^{\mathrm{i}m\hat{\varphi}}\otimes|L\rangle\langle L|$. By denoting the initial state of the joint system of probe and ancilla as $|\psi(l)\rangle\otimes|H\rangle$, a single query then leads the joint system to the state $|\Psi_{f}\rangle=\frac{1}{\sqrt{2}}\mathrm{e}^{-\mathrm{i}\alpha\hat{L}_{z}}(|\psi(l-m)\rangle\otimes|R\rangle+|\psi(l+m)\rangle\otimes|L\rangle)$ in the experiment.

For the single-query scenario, where the value of unknown parameter is fixed as $g=\pi/180$, a rotation angle of $\alpha=\ang{1}$ corresponds to the normalized evolution time of $T_{S}=1$. Meanwhile, the evolution time $T_{C}$ of the encoding process is represented by the order $m$ of the Q-plate ($T_{C}=m$), with $m=1$ corresponding to the normalized evolution time $T_{C}=1$. Here, we designate the dimensionless evolution time $T_{S}=T_{C}=T/2$. To estimate the unknown parameter from the final state, a practical projective measurement is devised for the experiment. As depicted in Fig.~\ref{fig:3}, a half-wave plate (HWP) followed by a $m$-order Q-plate are used to eliminate the topological charges on probe state. Theoretically, the HWP, with its optical axis set either parallel or perpendicular to the horizontal plane, flips the spin state of the photons. Consequently, the unitary evolution of this process, which comprises a HWP and an $m$-order Q-plate, can be represented as $\hat{U}_{m}=\mathrm{e}^{-\mathrm{i}m\hat{\varphi}}\otimes|L\rangle\langle L|+\mathrm{e}^{\mathrm{i}m\hat{\varphi}}\otimes|R\rangle\langle R|$. Subsequent to this, another HWP followed by a PBS are employed to project the final state of ancilla onto the $\ang{45}$ linearly polarized state $|+\rangle$ and the $-\ang{45}$ linearly polarized state $|-\rangle$. Given these configurations, the orthogonal projection operators performing on the final state can be represented as
\begin{equation}
	\label{eq:13}
	\begin{split}
		\hat{\Pi}_{+} &= \hat{U}_{m}^{\dagger}\left(\hat{\mathbb{I}}\otimes|+\rangle\langle+|\right)\hat{U}_{m}, \\
		\hat{\Pi}_{-} &= \hat{U}_{m}^{\dagger}\left(\hat{\mathbb{I}}\otimes|-\rangle\langle-|\right)\hat{U}_{m}.
	\end{split}
\end{equation}
Then we can calculate that
\begin{align}
	\hat{U}_{m}|\Psi_{f}\rangle &= \mathrm{e}^{-\mathrm{i}\alpha \hat{L}_{z}}|\psi(l)\rangle\otimes\frac{1}{\sqrt{2}}\left(\mathrm{e}^{\mathrm{i}m\alpha}|R\rangle+\mathrm{e}^{-\mathrm{i}m\alpha}|L\rangle\right) \nonumber \\
	&= \mathrm{e}^{-\mathrm{i}\frac{gT}{2}\hat{L}_{z}}|\psi(l)\rangle\otimes\frac{1}{\sqrt{2}}\left(\mathrm{e}^{\mathrm{i}gT^{2}/4}|R\rangle+\mathrm{e}^{-\mathrm{i}gT^{2}/4}|L\rangle\right), \label{eq:14}
\end{align}
wherein $T_{S}=\alpha/g=T/2$ and $T_{C}=m=T/2$. Thus, the projective measurement is equivalent to applying the operators $|+\rangle\langle+|$ and  $|-\rangle\langle-|$ to the final polarization state (ancilla) $|\phi_{f}\rangle=(\mathrm{e}^{\mathrm{i}gT^{2}/4}|R\rangle+\mathrm{e}^{-\mathrm{i}gT^{2}/4}|L\rangle)/\sqrt{2}$, which leads to the projective probabilities
\begin{equation}
	\label{eq:15}
	\begin{split}
		P_{+} &= \langle\Psi_{f}|\hat{\Pi}_{+}|\Psi_{f}\rangle = \frac{1}{2}\left[1+\sin\left(\frac{1}{2}gT^{2}\right)\right], \\
		P_{-} &= \langle\Psi_{f}|\hat{\Pi}_{-}|\Psi_{f}\rangle = \frac{1}{2}\left[1-\sin\left(\frac{1}{2}gT^{2}\right)\right].
	\end{split}
\end{equation}

Substituting $P_{+}$ and $P_{-}$ into Eq.~(\ref{eq:11}), the CFI of estimating $g$ from the projective measurement after a single query of the ITD encoding process and the parameterizing process is given by
\begin{equation}
	F(g) = T^{4}\frac{\cos^{2}\left(gT^{2}/2\right)}{4-4\sin^{2}\left(gT^{2}/2\right)} = \frac{1}{4}T^{4}. \label{eq:16}
\end{equation}
Then drawing on the CCRB, the precision limit in Eq.~(\ref{eq:6}) for estimating the parameter $g$ in the experiment can be obtained.

For the multiple-queries scenario, the evolution time lengths are fixed as $T_{C}=T_{S}=1$ and $T=T_{C}+T_{S}=2$, which means the topological order of the Q-plate is set to $m=1$ and the unknown parameter $g$ is directly determined by the rotation angle $\alpha$ of the photon beam profile. Therefore, the rotation angle is fixed as $\alpha=gT_{S}=\ang{1}$. Following $N$ queries of the ITD encoding process and the parameterizing process, we denote the final joint state of the probe and ancilla as $|\Psi_{f}^{(N)}\rangle$. We use experimental settings similar to those we've mentioned above to perform projective measurements on this final state, where an $N$-order Q-plate is employed in this case. Then the corresponding projective operators can be expressed as $\hat{\Pi}_{+}^{(N)}=\hat{U}_{N}^{\dagger}\left(\hat{\mathbb{I}}\otimes|+\rangle\langle+|\right)\hat{U}_{N}$ and $\hat{\Pi}_{-}^{(N)}=\hat{U}_{N}^{\dagger}\left(\hat{\mathbb{I}}\otimes|-\rangle\langle-|\right)\hat{U}_{N}$, where $\hat{U}_{N}=\mathrm{e}^{-\mathrm{i}N\hat{\varphi}}\otimes|L\rangle\langle L|+\mathrm{e}^{\mathrm{i}N\hat{\varphi}}\otimes|R\rangle\langle R|$. Similarly, we can calculate that
\begin{equation}
	\hat{U}_{N}|\Psi_{f}^{(N)}\rangle = \mathrm{e}^{-\mathrm{i}N\alpha\hat{L}_{z}}|\psi(l)\rangle\otimes\frac{1}{\sqrt{2}}\left(\mathrm{e}^{\mathrm{i}\frac{N^{2}+N}{2}\alpha}|R\rangle+\mathrm{e}^{-\mathrm{i}\frac{N^{2}+N}{2}\alpha}|L\rangle\right), \label{eq:17}
\end{equation}
wherein $\alpha=g$. The corresponding projective probabilities are then derived as
\begin{equation}
	\label{eq:18}
	\begin{split}
		P_{+}^{(N)} &= \langle\Psi_{f}^{(N)}|\hat{\Pi}_{+}^{(N)}|\Psi_{f}^{(N)}\rangle = \frac{1}{2}\left\{1+\sin\left[(N^{2}+N)\alpha\right]\right\}, \\
		P_{-}^{(N)} &= \langle\Psi_{f}^{(N)}|\hat{\Pi}_{-}^{(N)}|\Psi_{f}^{(N)}\rangle = \frac{1}{2}\left\{1-\sin\left[(N^{2}+N)\alpha\right]\right\},
	\end{split}
\end{equation}

Substituting $P_{+}^{(N)}$ and $P_{-}^{(N)}$ into Eq.~(\ref{eq:11}), the CFI of estimating $g$ from the projective measurement after $N$ queries of the ITD encoding process and the parameterizing process is given by
\begin{equation}
	F(\alpha) = (N^{2}+N)^{2}\frac{\cos^{2}\left[(N^{2}+N)^{2}\alpha\right]}{1-\sin^{2}\left[(N^{2}+N)^{2}\alpha\right]} = (N^{2}+N)^{2}. \label{eq:19}
\end{equation}
Then drawing on the CCRB, the precision limit in Eq.~(\ref{eq:7}) for estimating the parameter $\alpha$ ($g$) in the experiments can be obtained.

\subsection{Calculating RMSE from experimental data}
In the experiments, the estimated value $\tilde{g}$ of the parameter $g$ can be obtained from the detected photons number $\tilde{m}_{+}$ under the projection $\hat{\Pi}_{+}$ and the detected photons number $\tilde{m}_{-}$ under the projection $\hat{\Pi}_{-}$. To derive the expression of the estimator $\tilde{g}$ from the measurement results $\tilde{m}_{+}$ and $\tilde{m}_{-}$, we first calculate the log-likelihood function for the unknown parameter $g$ as
\begin{align}
	\ell(g|\{\tilde{m}_{+},\tilde{m}_{-}\}) &= \ln\left(\mathcal{P}_{0}\prod_{i=+,-}P_{i}^{\tilde{m}_{i}}\right) \nonumber \\
	&= \ln\mathcal{P}_{0}+\sum_{i=+,-}\tilde{m}_{i}P_{i}, \label{eq:20}
\end{align}
where
\begin{equation}
	\ln\mathcal{P}_{0} = \frac{\left(\sum_{i}\tilde{m}_{i}\right)!}{\prod_{i}\tilde{m}_{i}!}, \label{eq:21}
\end{equation}
accounts for all possible permutations. Then, by solving the likelihood equation
\begin{equation}
	\frac{\partial}{\partial g}\ell(g|\{\tilde{m}_{+},\tilde{m}_{-}\}) = 0, \label{eq:22}
\end{equation}
we can obtain the estimated value $\tilde{g}$ from measured data $\tilde{m}_{+}$ and $\tilde{m}_{-}$. The RMSE of estimating the parameter $g$ in the experiments can be calculated from
\begin{equation}
	\mathrm{RMSE}(g) = \sqrt{\frac{1}{L}\sum_{i}\left(\tilde{g}^{(i)}-g_{0}\right)}, \label{eq:23}
\end{equation}
where $g_{0}$ is the true value of parameter $g$ we set in the experiments, $L$ is the number of the trials repeated (which is 30 in our experiments), and $\tilde{g}^{(i)}$ is the estimated value of $i$-th measurement.

Furthermore, by substituting $P_{+}$ and $P_{-}$ from Eq.~(\ref{eq:15}) into the likelihood equation, we can calculate the estimator of the parameter $g$ with respect to various time lengths $T$ of a single-shot evolution, which comprises a ITD encoding process and a parameterizing process, in our experiments as
\begin{equation}
	\tilde{g}_{\mathrm{exp}} = \frac{1}{2T^{2}}\arcsin\left(\frac{\tilde{m}_{+}-\tilde{m}_{-}}{\tilde{m}_{+}+\tilde{m}_{-}}\right). \label{eq:24}
\end{equation}
Next, by substituting $P_{+}^{(N)}$ and $P_{-}^{(N)}$ from Eq.~(\ref{eq:18}) into the likelihood equation, we can calculate the estimator of the parameter $g$ (which equals the rotation angle $\alpha$ in this setting) with respect to various numbers $N$ of sequential evolutions in our experiments as
\begin{equation}
	\tilde{g}_{\mathrm{exp}}^{(N)} = \tilde{\alpha}_{\mathrm{exp}} = \frac{1}{N^{2}+N}\arcsin\left(\frac{\tilde{m}_{+}-\tilde{m}_{-}}{\tilde{m}_{+}+\tilde{m}_{-}}\right). \label{eq:25}
\end{equation}


\begin{acknowledgments}
	This work was supported by the National Natural Science Foundation of China (No. 62471289), Quantum Science and Technology-National Science and Technology Major Project (No.2021ZD0300703), Shanghai Municipal Science and Technology Major Project (Grant No. 2019SHZDZX01), the National Natural Science Foundation of China for the Excellent Young Scientists Fund (Hong Kong and Macau) Project 12322516, Ministry of Science and Technology, China (MOST2030) with Grant No. 2023200300600 and the Hong Kong Research Grant Council (RGC) through grant 27310822 and grant 17302724.
\end{acknowledgments}

\noindent {\bf Data and Materials Availability} All data needed to evaluate the conclusions in the paper are present in the paper and the Supplementary Materials and dataset \cite{Binke2026}.

\noindent {\bf Competing Interests} All authors declare that they have no competing interests.

\noindent {\bf Author Contribution} G.Z. conceived the research project, J.H. designed the scheme, B.X. constructed the theoretical model and carried out the experiments with assistance from J.H. and Y.Y., B.X. analyzed the data. B.X., J.H. and Y.Y. wrote the manuscript. All authors have read and approved the final version of the manuscript.

\noindent {\bf Corresponding Authors:} Jingzheng Huang and Yuxiang Yang.

\bibliography{bibliography}

\end{document}



\title{Supplementary Materials for Scaling Enhancement in Quantum Metrology via Indefinite-Time-Direction Encoding}


\author{Binke Xia}
\affiliation{State Key Laboratory of Photonics and Communications, Institute for Quantum Sensing and Information Processing, School of Automation and Intelligent Sensing, Shanghai Jiao Tong University, Shanghai 200240, China}
\affiliation{QICI Quantum Information and Computation Initiative, School of Computing and Data Science, The University of Hong Kong, Pokfulam Road, Hong Kong, China}
\author{Jingzheng Huang}
\email{jzhuang1983@sjtu.edu.cn}
\affiliation{State Key Laboratory of Photonics and Communications, Institute for Quantum Sensing and Information Processing, School of Automation and Intelligent Sensing, Shanghai Jiao Tong University, Shanghai 200240, China}
\affiliation{Hefei National Laboratory, Hefei 230088, China}
\affiliation{Shanghai Research Center for Quantum Sciences, Shanghai 201315, China}
\author{Yuxiang Yang}
\email{yxyang@hku.hk}
\affiliation{QICI Quantum Information and Computation Initiative, School of Computing and Data Science, The University of Hong Kong, Pokfulam Road, Hong Kong, China}
\author{Guihua Zeng}
\affiliation{State Key Laboratory of Photonics and Communications, Institute for Quantum Sensing and Information Processing, School of Automation and Intelligent Sensing, Shanghai Jiao Tong University, Shanghai 200240, China}
\affiliation{Hefei National Laboratory, Hefei 230088, China}
\affiliation{Shanghai Research Center for Quantum Sciences, Shanghai 201315, China}



\maketitle


\section{Ultimate Precision Limits in Quantum Metrology}
In this section, we calculate the quantum Fisher information (QFI) for both the standard scheme and our ITD-ending scheme. We further derive an upper bound on the QFI by employing the dynamical characteristic operator $\hat{V}_{S}=\partial_{g}\hat{H}_{S}(g)$ to reveal the dependence of the standard scheme on the probe resources. According to the quantum Cramér–Rao bound (QCRB) theory \cite{Helstrom_1973,PhysRevLett.72.3439}, the ultimate precision limits in quantum metrological schemes are determined by the inverse of the corresponding QFI.

\subsection{Calculating QFI}
In the generalized quantum metrological scheme with a unitary parameterizing process $\hat{U}(g)$, the unitary operator $\hat{U}(g)$ maps a quantum state $\hat{\rho}=\sum_{i}p_{i}|\eta_{i}\rangle\langle\eta_{i}|$ in the Hilbert space $\mathbb{H}$ to a parameterized state $\hat{\rho}(g)=\hat{U}(g)\hat{\rho}\hat{U}^{\dagger}(g)=\sum_{i}p_{i}\hat{U}(g)|\eta_{i}\rangle\langle\eta_{i}|\hat{U}^{\dagger}(g)$ in the projective Hilbert space (parameter space) $\mathbb{H}/U(1)$. In this case, the QFI in estimating $g$ from $\hat{\rho}(g)$ is given by \cite{Liu_2019}
\begin{equation}
	\mathcal{F}(g) = \sum_{i}4p_{i}\left(\langle\eta_{i}|\hat{\mathcal{H}}^{2}|\eta_{i}\rangle-\langle\eta_{i}|\hat{\mathcal{H}}|\eta_{i}\rangle^{2}\right)-\sum_{i\neq j}\frac{8p_{i}p_{j}}{p_{i}+p_{j}}\left|\langle\eta_{i}|\hat{\mathcal{H}}|\eta_{j}\rangle\right|^{2}, \label{eq:1}
\end{equation}
where
\begin{equation}
	\hat{\mathcal{H}} = \mathrm{i}\hat{U}^{\dagger}(g)\left[\frac{\partial\hat{U}(g)}{\partial g}\right], \label{eq:2}
\end{equation}
denotes the generator of parameter $g$. Theoretically, the ultimate precision limit in estimating $g$ is given by
\begin{equation}
	\delta g \ge \frac{1}{\sqrt{M\mathcal{F}(g)}}, \label{eq:3}
\end{equation}
where $M$ denoted the number of independent measurements.

\subsection{QFI in standard scheme}
In the standard quantum metrological scheme, a prepare probe state $\hat{\rho}_{S}=\sum_{i}p_{i}|\psi_{i}\rangle\langle\psi_{i}|$ evolves under the parameterizing process $\hat{U}_{S}(g)$ described by a parameter-dependent Hamiltonian $\hat{H}_{S}(g)$ during a time $T_{S}$. Then the generator of $g$ can be further derived as \cite{Pang_2017}
\begin{equation}
	\hat{\mathcal{H}}_{S}=\int_{0}^{T_{S}}\hat{U}_{S}^{\dagger}(0\to t)\left[\frac{\partial \hat{H}_{S}(g)}{\partial g}\right]\hat{U}_{S}(0\to t)\,\mathrm{d}t, \label{eq:4}
\end{equation}
where we denote the dynamical characteristic operator as
\begin{equation}
	\hat{V}_{S} = \frac{\partial \hat{H}_{S}(g)}{\partial g}, \label{eq:5}
\end{equation}
it reflects the $g$-dependence of the parameterizing dynamics. Substituting $\hat{\rho}_{S}$ and $\hat{H}_{S}$ into Eq.~(\ref{eq:1}), we can derive that the QFI of the standard scheme is upper bounded by
\begin{equation}
	\mathcal{F}_{S}(g) \le \sum_{i}4p_{i}\left(\langle\psi_{i}|\hat{\mathcal{H}}_{S}^{2}|\psi_{i}\rangle-\langle\psi_{i}|\hat{\mathcal{H}}_{S}|\psi_{i}\rangle^{2}\right) = 4\Delta\bar{\mathcal{H}}_{S}^{2}, \label{eq:6}
\end{equation}
since the eigenvalue $p_{i}$ in the spectrum decomposition of state $\hat{\rho}_{S}$ satisfies $p_{i} >0$. Here $\Delta\bar{\mathcal{H}}_{S}^{2}$ denotes the average variance of $\hat{\mathcal{H}}_{S}$ with respect to the probe state. It can be further denoted as
\begin{equation*}
	\Delta\bar{\mathcal{H}}_{S}^{2} = \sum_{i}p_{i}\left(\langle\psi_{i}|\hat{\mathcal{H}}_{S}^{2}|\psi_{i}\rangle-\langle\psi_{i}|\hat{\mathcal{H}}_{S}|\psi_{i}\rangle^{2}\right) = \sum_{i}p_{i}\Delta_{i}\mathcal{H}_{S}^{2},
\end{equation*}
where $\Delta_{i}\mathcal{H}_{S}^{2}=(\langle\psi_{i}|\hat{\mathcal{H}}_{S}^{2}|\psi_{i}\rangle-\langle\psi_{i}|\hat{\mathcal{H}}_{S}|\psi_{i}\rangle^{2})$ denotes the variance of $\hat{\mathcal{H}}_{S}$ on the $i$-th eigenstate $|\psi_{i}\rangle$ of the probe state. Substituting the expression of $\hat{\mathcal{H}}_{S}$ in Eq.~(\ref{eq:3}), we can then derive that (for simplicity, we replace $\hat{U}_{S}(0\to t)$ with $\hat{U}_{S}(t)$ hereafter)
\begin{align*}
	\Delta_{i}\mathcal{H}_{S}^{2} &= \langle\psi_{i}|\left(\int_{0}^{T_{S}}\hat{U}_{S}^{\dagger}(t)\hat{V}_{S}\hat{U}_{S}(t)\mathrm{d}t\right)^{2}|\psi_{i}\rangle-\left(\langle\psi_{i}|\int_{0}^{T_{S}}\hat{U}_{S}^{\dagger}(t)\hat{V}_{S}\hat{U}_{S}(t)\mathrm{d}t|\psi_{i}\rangle\right)^{2} \\
	&= \int_{0}^{T_{S}}\langle\psi_{i}|\hat{U}_{S}^{\dagger}(t)\hat{V}_{S}\hat{U}_{S}(t)\mathrm{d}t\cdot\left(\hat{\mathbb{I}}-|\psi_{i}\rangle\langle\psi_{i}|\right)\cdot\int_{0}^{T_{S}}\hat{U}_{S}^{\dagger}(t)\hat{V}_{S}\hat{U}_{S}(t)|\psi_{i}\rangle\mathrm{d}t \\
	&= \int_{0}^{T_{S}}\langle\psi_{i}|\hat{U}_{S}^{\dagger}(t)\hat{V}_{S}\hat{U}_{S}(t)\left(\hat{\mathbb{I}}-|\psi_{i}\rangle\langle\psi_{i}|\right)\mathrm{d}t\cdot\int_{0}^{T_{S}}\left(\hat{\mathbb{I}}-|\psi_{i}\rangle\langle\psi_{i}|\right)\hat{U}_{S}^{\dagger}(t)\hat{V}_{S}\hat{U}_{S}(t)|\psi_{i}\rangle\mathrm{d}t.
\end{align*}
Drawing on the Cauchy–Schwarz inequality of $|\int_{a}^{b}f(x)\mathrm{d}x|^{2}\le (b-a)\int_{a}^{b}|f(x)|^{2}\mathrm{d}x$, we can derive
\begin{align*}
	\Delta_{i}\mathcal{H}_{S}^{2} &\le T_{S}\int_{0}^{T_{S}}\langle\psi_{i}|\hat{U}_{S}^{\dagger}(t)\hat{V}_{S}\hat{U}_{S}(t)\left(\hat{\mathbb{I}}-|\psi_{i}\rangle\langle\psi_{i}|\right)\hat{U}_{S}^{\dagger}(t)\hat{V}_{S}\hat{U}_{S}(t)|\psi_{i}\rangle\mathrm{d}t \\
	&= T_{S}\int_{0}^{T_{S}}\left(\langle\psi_{i}|\hat{U}_{S}^{\dagger}(t)\hat{V}_{S}^{2}\hat{U}_{S}(t)|\psi_{i}\rangle-\langle\psi_{i}|\hat{U}_{S}^{\dagger}(t)\hat{V}_{S}\hat{U}_{S}(t)|\psi_{i}\rangle^{2}\right)\mathrm{d}t.
\end{align*}
Then we can derive
\begin{align}
	\Delta\bar{\mathcal{H}}_{S}^{2} &= \sum_{i}p_{i}\Delta_{i}\mathcal{H}_{S}^{2} \nonumber\\
	&\le T_{S}\int_{0}^{T_{S}}\sum_{i}p_{i}\left(\langle\psi_{i}|\hat{U}_{S}^{\dagger}(t)\hat{V}_{S}^{2}\hat{U}_{S}(t)|\psi_{i}\rangle-\langle\psi_{i}|\hat{U}_{S}^{\dagger}(t)\hat{V}_{S}\hat{U}_{S}(t)|\psi_{i}\rangle^{2}\right)\mathrm{d}t \nonumber\\
	&= T_{S}\int_{0}^{T_{S}}\left.\Delta\bar{V}_{S}^{2}\right|_{\hat{\rho}_{S}(t)}\,\mathrm{d}t, \label{eq:7}
\end{align}
where $\hat{\rho}_{S}(t)=\sum_{i}p_{i}\hat{U}_{S}(t)|\psi_{i}\rangle\langle\psi_{i}|\hat{U}_{S}^{\dagger}(t)$ is the probe state at time $t$. Therefore, the corresponding QFI is upper bounded by
\begin{equation}
	\mathcal{F}_{S}(g) \le 4T_{S}\int_{0}^{T_{S}}\left.\Delta\bar{V}_{S}^{2}\right|_{\hat{\rho}_{S}(t)}\,\mathrm{d}t, \label{eq:8}
\end{equation}
which is completely determined by the dynamical resource $\Delta\bar{V}_{S}$ on the probe state. Imposing the maximum resource constraint
\begin{equation*}
	\mathcal{V} = \max_{\hat{\rho}_{S}(t)} \Delta\bar{V}_{S} \le V,
\end{equation*}
the upper bound of QFI in the standard scheme is finally given by
\begin{equation}
	\mathcal{F}_{S}(g) \le 4V^{2}T_{S}^{2}. \label{eq:9}
\end{equation}
This result indicates that, when the maximum dynamical resource on the probe state is bounded, the ultimate precision limit in the standard scheme scales at most linearly with the evolution time $T=T_{S}$.

Next, we allow the probe to sequentially pass through the identical parameterizing process $\hat{U}_{S}(g)$ $N$ times within the standard quantum metrological scheme. Consequently, the total sequential evolution can be expressed as $\hat{U}_{S}^{(N)}=[\hat{U}_{S}(g)]^{N}$, the corresponding generator is then given by
\begin{equation*}
	\hat{\mathcal{H}}_{S}^{(N)} = \mathrm{i}\left[\hat{U}_{S}^{(N)}\right]^{\dagger}\left[\partial_{g}\hat{U}_{S}^{(N)}\right] = \mathrm{i}\sum_{j=0}^{N-1}\left(\hat{U}_{S}^{\dagger}\right)^{j}\hat{U}_{S}^{\dagger}\left(\partial_{g}\hat{U}_{S}\right)\left(\hat{U}_{S}\right)^{j} = \sum_{j=0}^{N-1}\left(\hat{U}_{S}^{\dagger}\right)^{j}\hat{\mathcal{H}}_{S}\left(\hat{U}_{S}\right)^{j}.
\end{equation*}
Similarly, we can calculate that
\begin{equation*}
	\left[\Delta\bar{\mathcal{H}}_{S}^{(N)}\right]^{2} = \sum_{i}p_{i}\left(\langle\psi_{i}|[\hat{\mathcal{H}}_{S}^{(N)}]^{2}|\psi_{i}\rangle-\langle\psi_{i}|\hat{\mathcal{H}}_{S}^{(N)}|\psi_{i}\rangle^{2}\right) = \sum_{i}p_{i}\left[\Delta_{i}\mathcal{H}_{S}^{(N)}\right]^{2},
\end{equation*}
where $[\Delta_{i}\mathcal{H}_{S}^{(N)}]^{2}$ denotes the variance of $\hat{\mathcal{H}}_{S}^{(N)}$ on the $i$-th eigenstate $|\psi_{i}\rangle$ of the probe state, and
\begin{align*}
	\left[\Delta_{i}\mathcal{H}_{S}^{(N)}\right]^{2} &= \langle\psi_{i}|\left(\sum_{j=0}^{N-1}\left(\hat{U}_{S}^{\dagger}\right)^{j}\hat{\mathcal{H}}_{S}\left(\hat{U}_{S}\right)^{j}\right)^{2}|\psi_{i}\rangle-\left(\langle\psi_{i}|\sum_{j=0}^{N-1}\left(\hat{U}_{S}^{\dagger}\right)^{j}\hat{\mathcal{H}}_{S}\left(\hat{U}_{S}\right)^{i}|\psi_{j}\rangle\right)^{2} \\
	&= \sum_{j=0}^{N-1}\langle\psi_{i}|\left(\hat{U}_{S}^{\dagger}\right)^{j}\hat{\mathcal{H}}_{S}\left(\hat{U}_{S}\right)^{j}\cdot\left(\hat{\mathbb{I}}-|\psi_{i}\rangle\langle\psi_{i}|\right)\cdot\sum_{j=0}^{N-1}\left(\hat{U}_{S}^{\dagger}\right)^{j}\hat{\mathcal{H}}_{S}\left(\hat{U}_{S}\right)^{j}|\psi_{i}\rangle \\
	&= \sum_{j=0}^{N-1}\langle\psi_{i}|\left(\hat{U}_{S}^{\dagger}\right)^{j}\hat{\mathcal{H}}_{S}\left(\hat{U}_{S}\right)^{j}\left(\hat{\mathbb{I}}-|\psi_{i}\rangle\langle\psi_{i}|\right)\cdot\sum_{j=0}^{N-1}\left(\hat{\mathbb{I}}-|\psi_{i}\rangle\langle\psi_{i}|\right)\left(\hat{U}_{S}^{\dagger}\right)^{j}\hat{\mathcal{H}}_{S}\left(\hat{U}_{S}\right)^{j}|\psi_{i}\rangle.
\end{align*}
Drawing on the Cauchy–Schwarz inequality of $|\sum_{j}x_{j}|^{2}\le n\sum_{j}|x_{j}|^{2}$, where $n$ denotes the number of terms in the summation, we can derive
\begin{align*}
	\left[\Delta_{i}\mathcal{H}_{S}^{(N)}\right]^{2} &\le N\sum_{j=0}^{N-1}\langle\psi_{i}|\left(\hat{U}_{S}^{\dagger}\right)^{j}\hat{\mathcal{H}}_{S}\left(\hat{U}_{S}\right)^{j}\left(\hat{\mathbb{I}}-|\psi_{i}\rangle\langle\psi_{i}|\right)\left(\hat{U}_{S}^{\dagger}\right)^{j}\hat{\mathcal{H}}_{S}\left(\hat{U}_{S}\right)^{j}|\psi_{i}\rangle \\
	&= N\sum_{j=0}^{N-1}\left[\langle\psi_{i}|\left(\hat{U}_{S}^{\dagger}\right)^{j}\hat{\mathcal{H}}_{S}^{2}\left(\hat{U}_{S}\right)^{j}|\psi_{i}\rangle-\langle\psi_{i}|\left(\hat{U}_{S}^{\dagger}\right)^{j}\hat{\mathcal{H}}_{S}\left(\hat{U}_{S}\right)^{j}|\psi_{i}\rangle^{2}\right].
\end{align*}
Then we can derive that
\begin{align*}
	\left[\Delta\bar{\mathcal{H}}_{S}^{(N)}\right]^{2} &= \sum_{i}p_{i}\left[\Delta_{i}\mathcal{H}_{S}^{(N)}\right]^{2} \\
	&\le N\sum_{j=0}^{N-1}\sum_{i}p_{i}\left[\langle\psi_{i}|\left(\hat{U}_{S}^{\dagger}\right)^{j}\hat{\mathcal{H}}_{S}^{2}\left(\hat{U}_{S}\right)^{j}|\psi_{i}\rangle-\langle\psi_{i}|\left(\hat{U}_{S}^{\dagger}\right)^{j}\hat{\mathcal{H}}_{S}\left(\hat{U}_{S}\right)^{j}|\psi_{i}\rangle^{2}\right] \\
	&= N\sum_{j=1}^{N}\left.\Delta\bar{\mathcal{H}}_{S}^{2}\right|_{\hat{\rho}_{S}(j)},
\end{align*}
where $\hat{\rho}_{S}(j)=\sum_{i}p_{i}(\hat{U}_{S})^{j-1}|\psi_{i}\rangle\langle\psi_{i}|(\hat{U}_{S}^{\dagger})^{j-1}$ denotes the input probe state of the $j$-th parameterizing process. Substituting Eq.~(\ref{eq:7}) into this result, we can further derive that
\begin{equation*}
	\left[\Delta\bar{\mathcal{H}}_{S}^{(N)}\right]^{2} \le N\sum_{j=1}^{N}T_{S}\int_{0}^{T_{S}}\left.\Delta\bar{V}_{S}^{2}\right|_{\hat{\rho}_{S}(j,t)}\,\mathrm{d}t,
\end{equation*}
where $\hat{\rho}_{S}(j,t)=\sum_{i}p_{i}\hat{U}_{S}(t)(\hat{U}_{S})^{j-1}|\psi_{i}\rangle\langle\psi_{i}|(\hat{U}_{S}^{\dagger})^{j-1}\hat{U}_{S}^{\dagger}(t)$ denotes the evolved probe state at time $t$ in the $j$-th parameterizing process. Considering the maximum resource constraint on the probe state
\begin{equation*}
	\mathcal{V} = \max_{\hat{\rho}_{S}(j,t)} \Delta\bar{V}_{S} \le V,
\end{equation*}
the upper bound of QFI in the standard scheme is finally given by
\begin{equation}
	\mathcal{F}_{S}^{(N)}(g) \le 4V^{2}N^{2}T_{S}^{2}. \label{eq:10}
\end{equation}
Substituting this result into QCRB, we can finally give the linear-scaling precision limit as shown in Eq.~(3) of the main text.

\subsection{QFI in ITD-encoding scheme}
Next, we calculate the QFI in our ITD-encoding scheme. The evolution of ITD encoding process is given by $\hat{U}_{I}=\hat{U}_{C}\otimes|0\rangle\langle 0|+\hat{U}_{C}^{\dagger}\otimes|1\rangle\langle 1|$, where $|0\rangle$ and $|1\rangle$ are the eigenstates of a two-level ancilla and $\hat{U}_{C}=\exp(-\mathrm{i}\hat{H}_{C}T_{C})$ denotes the evolution of encoding process with a definite time direction and $T_{C}$ is its duration. In our scheme, the encoding Hamiltonian satisfies $[\hat{H}_{C},\hat{V}_{S}]=\mathrm{i}$ and the commutator $[\hat{H}_{C},\hat{H}_{S}]$ commutes with both $\hat{H}_{C}$ and $\hat{H}_{S}$ at any time, i.e. $[\hat{H}_{C},[\hat{H}_{C},\hat{H}_{S}(t_{\mu})]]=[\hat{H}_{S}(t_{\nu}),[\hat{H}_{C},\hat{H}_{S}(t_{\mu})]]=0$ for any time $t_{\mu}$ and $t_{\nu}$ during the parameterizing process. By taking the partial derivative with respect to $g$ on both sides of the equation $[\hat{H}_{S}(t_{\nu}),[\hat{H}_{C},\hat{H}_{S}(t_{\mu})]]=0$, we can calculate that
\begin{equation*}
	[\hat{V}_{S}(t_{\nu}),[\hat{H}_{C},\hat{H}_{S}(t_{\mu})]]+[\hat{H}_{S}(t_{\nu}),[\hat{H}_{C},\hat{V}_{S}(t_{\mu})]] = 0.
\end{equation*}
Given that the Hamiltonian $\hat{H}_{C}$ satisfies $[\hat{H}_{C},\hat{V}_{S}]=\mathrm{i}$, we can derive that
\begin{equation*}
	[\hat{V}_{S}(t_{\nu}),[\hat{H}_{C},\hat{H}_{S}(t_{\mu})]] = 0,
\end{equation*}
which implies that the commutator $[\hat{H}_{C},\hat{H}_{S}]$ also commutes with operator $\hat{V}_{S}$ at any time.

Given that the entire evolution comprising a ITD encoding process and a parameterizing process is written as $\hat{U}_{IS}=\hat{U}_{S}\hat{U}_{C}\otimes|0\rangle\langle 0|+\hat{U}_{S}\hat{U}_{C}^{\dagger}\otimes|1\rangle\langle 1|$, the generator of parameter $g$ in our scheme given by
\begin{equation}
	\hat{\mathcal{H}}_{I} = \mathrm{i}\hat{U}_{IS}^{\dagger}(g)[\partial_{g}\hat{U}_{IS}(g)] = \hat{U}_{C}^{\dagger}\hat{\mathcal{H}}_{S}\hat{U}_{C}\otimes|0\rangle\langle 0|+\hat{U}_{C}\hat{\mathcal{H}}_{S}\hat{U}_{C}^{\dagger}\otimes|1\rangle\langle 1|, \label{eq:11}
\end{equation}
where
\begin{equation*}
	\hat{U}_{C}^{\dagger}\hat{\mathcal{H}}_{S}\hat{U}_{C} = \int_{0}^{T_{S}}\hat{U}_{C}^{\dagger}\hat{U}_{S}^{\dagger}(t)\hat{V}_{S}\hat{U}_{S}(t)\hat{U}_{C}\,\mathrm{d}t = \int_{0}^{T_{S}}\hat{U}_{S}^{\dagger}(t)\hat{U}_{C}^{\dagger}\hat{V}_{S}\hat{U}_{C}\hat{U}_{S}(t)\,\mathrm{d}t,
\end{equation*}
since the commutator $[\hat{H}_{C},\hat{H}_{S}]$ commutes with operators $\hat{H}_{C}$, $\hat{H}_{S}$ and $\hat{V}_{S}$ at any time. Given that $[\hat{H}_{C},\hat{V}_{S}]=\mathrm{i}$, we can calculate that
\begin{equation*}
	\hat{U}_{C}^{\dagger}\hat{V}_{S}\hat{U}_{C} = \mathrm{e}^{\mathrm{i}\hat{H}_{C}T_{C}}\,\hat{V}_{S}\,\mathrm{e}^{-\mathrm{i}\hat{H}_{C}T_{C}} = \hat{V}_{S}-T_{C},
\end{equation*}
which leads to
\begin{equation}
	\hat{U}_{C}^{\dagger}\hat{\mathcal{H}}_{S}\hat{U}_{C} = \hat{\mathcal{H}}_{S}-T_{C}T_{S}. \label{eq:12}
\end{equation}
Similarly, we can derive that
\begin{equation}
	\hat{U}_{C}\hat{\mathcal{H}}_{S}\hat{U}_{C}^{\dagger} = \hat{\mathcal{H}}_{S}+T_{C}T_{S}. \label{eq:13}
\end{equation}
Substituting Eqs. (\ref{eq:12}) and (\ref{eq:13}) into Eq.~(\ref{eq:11}), the generator in our scheme is finally given by
\begin{equation}
	\hat{\mathcal{H}}_{I} = \hat{\mathcal{H}}_{S}\otimes\hat{\mathbb{I}}-T_{C}T_{S}\otimes\hat{\sigma}_{z}, \label{eq:14}
\end{equation}
where $\hat{\sigma}_{z}=|0\rangle\langle 0|-|1\rangle\langle 1|$ is the Pauli-Z operator acting on the ancilla.

Since the ancilla in our scheme is a pure state $\hat{\rho}_{A}=|\phi\rangle\langle\phi|$, we denote the initial state of the joint system of the probe and ancilla as $\hat{\rho}_{SA}=\hat{\rho}_{S}\otimes\hat{\rho}_{A}=\sum_{i}p_{i}|\psi_{i}\rangle\langle\psi_{i}|\otimes|\phi\rangle\langle\phi|=\sum_{i}p_{i}|\Psi_{i}\rangle\langle\Psi_{i}|$, where $|\Psi_{i}\rangle=|\psi_{i}\rangle\otimes|\phi\rangle$ is the eigenstates of the spectrum decomposition of the joint state. Substituting the initial state $\hat{\rho}_{SA}$ and the generator $\hat{H}_{I}$ into Eq.~(\ref{eq:1}), the QFI of our ITD-encoding scheme is given by
\begin{align*}
	\mathcal{F}_{I}(g) &= \sum_{i}4p_{i}\left(\langle\Psi_{i}|\hat{\mathcal{H}}_{I}^{2}|\Psi_{i}\rangle-\langle\Psi_{i}|\hat{\mathcal{H}}_{I}|\Psi_{i}\rangle^{2}\right)-\sum_{i\neq j}\frac{8p_{i}p_{j}}{p_{i}+p_{j}}\left|\langle\Psi_{i}|\hat{\mathcal{H}}_{I}|\Psi_{j}\rangle\right|^{2} \\
	&= \sum_{i}4p_{i}\left[\langle\psi_{i}|\hat{\mathcal{H}}_{S}^{2}|\psi_{i}\rangle+T_{C}^{2}T_{S}^{2}-2T_{C}T_{S}\langle\psi_{i}|\hat{\mathcal{H}}_{S}|\psi_{i}\rangle\langle\phi|\hat{\sigma}_{z}|\phi\rangle-\left(\langle\psi_{i}|\hat{\mathcal{H}}_{S}|\psi_{i}\rangle-T_{C}T_{S}\langle\phi|\hat{\sigma}_{z}|\phi\rangle\right)^{2}\right] \\
	&\qquad-\sum_{i\neq j}\frac{8p_{i}p_{j}}{p_{i}+p_{j}}\left|\langle\Psi_{i}|\hat{\mathcal{H}}_{S}|\Psi_{j}\rangle\right|^{2} \\
	=& \sum_{i}4p_{i}\left(\langle\psi_{i}|\hat{\mathcal{H}}_{S}^{2}|\psi_{i}\rangle-\langle\psi_{i}|\hat{\mathcal{H}}_{S}|\psi_{i}\rangle^{2}\right)-\sum_{i\neq j}\frac{8p_{i}p_{j}}{p_{i}+p_{j}}\left|\langle\Psi_{i}|\hat{\mathcal{H}}_{S}|\Psi_{j}\rangle\right|^{2}+4T_{C}^{2}T_{S}^{2}\left(1-\langle\phi|\hat{\sigma}_{z}|\phi\rangle^{2}\right).
\end{align*}
Therefore, the QFI of our scheme is given by a summation of the QFI $\mathcal{F}_{S}(g)$ of the standard scheme and a $T^{4}$-scaling increment
\begin{equation}
	\mathcal{F}_{I}(g) = \mathcal{F}_{S}(g)+4T_{C}^{2}T_{S}^{2}\left(1-\bar{\sigma}_{z}^{2}\right), \label{eq:15}
\end{equation}
$\bar{\sigma}_{z}=\langle\phi|\hat{\sigma}_{z}|\phi\rangle$ denotes the average value of Pauli operator $\hat{\sigma}_{z}$ on the ancilla $|\phi\rangle$.

Subsequently, we allow both the probe and the ancilla to sequentially pass through the identical metrological process $\hat{U}_{IS}$, which comprises an ITD encoding process $\hat{U}_{I}$ prior to the parameterizing process $\hat{U}_{S}(g)$, $N$ times within our quantum metrological scheme. Consequently, the total sequential evolution of the joint system of the probe and ancilla can be expressed as $\hat{U}_{IS}^{(N)}=(\hat{U}_{S}\hat{U}_{C})^{N}\otimes|0\rangle\langle 0|+(\hat{U}_{S}\hat{U}_{C}^{\dagger})^{N}\otimes|1\rangle\langle 1|$, the corresponding generator is then given by
\begin{align*}
	\hat{\mathcal{H}}_{I}^{(N)} &= \mathrm{i}\left[\hat{U}_{IS}^{(N)}\right]^{\dagger}\left[\partial_{g}\hat{U}_{IS}^{(N)}\right] \\
	&= \mathrm{i}\left(\hat{U}_{C}^{\dagger}\hat{U}_{S}^{\dagger}\right)^{N}\sum_{j=0}^{N-1}\left(\hat{U}_{S}\hat{U}_{C}\right)^{N-1-j}\left(\partial_{g}\hat{U}_{S}\right)\hat{U}_{C}\left(\hat{U}_{S}\hat{U}_{C}\right)^{j}\otimes|0\rangle\langle 0| \\
	&\qquad+\mathrm{i}\left(\hat{U}_{C}\hat{U}_{S}^{\dagger}\right)^{N}\sum_{j=0}^{N-1}\left(\hat{U}_{S}\hat{U}_{C}^{\dagger}\right)^{N-1-j}\left(\partial_{g}\hat{U}_{S}\right)\hat{U}_{C}^{\dagger}\left(\hat{U}_{S}\hat{U}_{C}^{\dagger}\right)^{j}\otimes|1\rangle\langle 1| \\
	&= \sum_{j=0}^{N-1}\left(\hat{U}_{C}^{\dagger}\hat{U}_{S}^{\dagger}\right)^{j}\hat{U}_{C}^{\dagger}\hat{\mathcal{H}}_{S}\hat{U}_{C}\left(\hat{U}_{S}\hat{U}_{C}\right)^{j}\otimes|0\rangle\langle 0|+\sum_{j=0}^{N-1}\left(\hat{U}_{C}\hat{U}_{S}^{\dagger}\right)^{j}\hat{U}_{C}\hat{\mathcal{H}}_{S}\hat{U}_{C}^{\dagger}\left(\hat{U}_{S}\hat{U}_{C}^{\dagger}\right)^{j}\otimes|1\rangle\langle 1|.
\end{align*}
Since the commutator $[\hat{H}_{C},\hat{H}_{S}]$ commutes with the operators $\hat{H}_{C}$, $\hat{H}_{S}$ and $\hat{V}_{S}$ at any time, it also commutes with the generator $\hat{\mathcal{H}}_{S}=\int_{0}^{T_{S}}\hat{U}_{S}^{\dagger}(0\to t)\hat{V}_{S}\hat{U}_{S}(0\to t)\mathrm{d}t$. Therefore, we can further derive $\hat{\mathcal{H}}_{I}^{(N)}$ as
\begin{align*}
	\hat{\mathcal{H}}_{I}^{(N)} &= \sum_{j=0}^{N-1}\left(\hat{U}_{S}^{\dagger}\right)^{j}\left(\hat{U}_{C}^{\dagger}\right)^{j+1}\hat{\mathcal{H}}_{S}\left(\hat{U}_{C}\right)^{j+1}\left(\hat{U}_{S}\right)^{j}\otimes|0\rangle\langle 0| \\
	&\qquad+\sum_{j=0}^{N-1}\left(\hat{U}_{S}^{\dagger}\right)^{j}\left(\hat{U}_{C}\right)^{j+1}\hat{\mathcal{H}}_{S}\left(\hat{U}_{C}^{\dagger}\right)^{j+1}\left(\hat{U}_{S}\right)^{j}\otimes|1\rangle\langle 1|.
\end{align*}
Combining with Eqs. (\ref{eq:12}) and (\ref{eq:13}), we can finally calculate that
\begin{align}
	\hat{\mathcal{H}}_{I}^{(N)} &= \sum_{j=0}^{N-1}\left(\hat{U}_{S}^{\dagger}\right)^{j}\hat{\mathcal{H}}_{S}\left(\hat{U}_{S}\right)^{j}\otimes\hat{\mathbb{I}}-\sum_{j=1}^{N}j\,T_{C}T_{S}\otimes\hat{\sigma}_{z} \nonumber\\
	&= \hat{\mathcal{H}}_{S}^{(N)}\otimes\hat{\mathbb{I}}-\frac{N^{2}+N}{2}T_{C}T_{S}\otimes\hat{\sigma}_{z}. \label{eq:16}
\end{align}
Substituting $\hat{\rho}_{SA}$ and $\hat{\mathcal{H}}_{I}^{(N)}$ into Eq.~(\ref{eq:1}), we can calculate the QFI of our scheme with $N$ queries as
\begin{align*}
	\mathcal{F}_{I}^{(N)}(g) &= \sum_{i}4p_{i}\left(\langle\psi_{i}|[\hat{\mathcal{H}}_{S}^{(N)}]^{2}|\psi_{i}\rangle-\langle\psi_{i}|\hat{\mathcal{H}}_{S}^{(N)}|\psi_{i}\rangle^{2}\right)-\sum_{i\neq j}\frac{8p_{i}p_{j}}{p_{i}+p_{j}}\left|\langle\Psi_{i}|\hat{\mathcal{H}}_{S}^{(N)}|\Psi_{j}\rangle\right|^{2} \\
	&\qquad+(N^{2}+N)^{2}T_{C}^{2}T_{S}^{2}\left(1-\langle\phi|\hat{\sigma}_{z}|\phi\rangle^{2}\right),
\end{align*}
which leads to the result
\begin{equation}
	\mathcal{F}_{I}^{(N)}(g) = \mathcal{F}_{S}^{(N)}(g)+T_{C}^{2}T_{S}^{2}\left(N^{2}+N\right)^{2}\left(1-\bar{\sigma}_{z}^{2}\right), \label{eq:17}
\end{equation}
as shown in Eq.~(4) of the main text. This result indicates that the nonlinear-scaling enhancement of the precision in our scheme does not rely any information intrinsic to the probes themselves.

\section{Algorithm of calculating coincidence counts}
\begin{algorithm}[b]
	\caption{Calculation of coincidence counts}
	\label{alg:1}
	\tcc{Assume that SPD A detects the herald photons and SPD B detects the signal photons. For a single-photon pair, the herald photon should be detected by SPD A before the signal photon is detected by SPD B.}
	\KwIn{Timestamps from SPD A and SPD B: $\mathcal{T}_{A}=\left\{t_{1}^{(A)},t_{2}^{(A)},\cdots,t_{m}^{(A)}\right\}$, $\mathcal{T}_{B}=\left\{t_{1}^{(B)},t_{2}^{(B)},\cdots,t_{n}^{(B)}\right\}$.}
	\KwOut{Coincidence counts $\mathcal{C}=\left\{c_{1},c_{2},\cdots,c_{k}\right\}$ between SPD A and SPD B; Corresponding delay times $T=\left\{t_{1}, t_{2}, \cdots, t_{k}\right\}$ of signal photons.}
	Initialize $T$ \tcc*{The step size is set to $\SI{5}{\ps}$ and the span ranges from $\SI{1}{\ns}$ to $\SI{50}{\ns}$ in our experiments.}
	Initialize $\mathcal{C}$ \tcc*{The values of coincidence counts are all set to 0.}
	Define $\mathcal{T}_{I}=\left\{t_{1}^{(I)},t_{2}^{(I)},\cdots,t_{n}^{(I)}\right\}$ \tcc*{$\mathcal{T}_{I}$ is the set of the time intervals between every detected single-photon pair.}
	Set $w$ \tcc*{$w$ is the gate width for coincidence counting, which is set to $\SI{2}{\ns}$ in our experiments.}
	~\\
	\tcc{Search the time intervals between every detected single-photon pair.}
	\For{$i\leftarrow 1$ \KwTo $n$}{\For{$j\leftarrow 1$ \KwTo $m$}{\eIf{$t_{i}^{(B)}-t_{j}^{(A)}>0$}{$t_{i}^{(I)}\leftarrow t_{i}^{(B)}-t_{j}^{(A)}$}{Break to the outter for loop}}}
	~\\
	\tcc{Calculate the coincidence counts with different delay times in $T$}
	\For{$i\leftarrow 1$ \KwTo $k$}{\For{$j\leftarrow 1$ \KwTo $n$}{\If{$t_{j}^{(I)}>t_{i}-w/2$ \textbf{and} $t_{j}^{(I)}<t_{i}+w/2$}{$c_{i}=c_{i}+1$}}}
\end{algorithm}

In our experiments, we recorded the timestamps of the detected photons for each single-photon detector (SPD) using the Moku:Pro device provided by Liquid Instruments, which offers a time jitter of less than $\SI{20}{\ps}$. We supplemented the original timestamp data detected by the three SPDs with eight individual files. These files correspond to an evolution time length $T=T_{S}+T_{C}$ of 2, 4, 6, and 8 for a single-shot evolution ($N=1$), and an iteration numbers $N$ of 1, 2, 3, and 4 for a fixed time length ($T=2$), respectively. Each file contains three columns of timestamp data, corresponding to the timestamps of pulse signals outputted by SPD 1, SPD 2, and SPD 3, respectively. Subsequently, we calculated the coincidence counts between SPD 2 and SPD 1, as well as between SPD 3 and SPD 1, using the timestamp data in each file.

In this work, we developed an algorithm for calculating the coincidence counts between the timestamp data of two different SPDs. This algorithm searches the time intervals between every detected single-photon pair and then determines the coincidence counts with various delay times of the signal photons. This method offers lower computational complexity and reduced computation time compared to the traditional approach of shifting the timestamps of signal photons to calculate the coincidence counts. The detailed algorithm is presented in Algorithm \ref{alg:1}.

Based on this algorithm, we calculated the coincidence counts between SPD 2 and SPD 1, as well as between SPD 3 and SPD 1, with varying delay times of the signal photons. In our calculations, the delay times range from $\SI{1}{\ns}$ to $\SI{50}{\ns}$, with a step size of $\SI{5}{\ps}$. Considering the time jitter in SPDs, the gate width for coincidence counting in our calculations is set to $\SI{2}{\ns}$. Subsequently, the number of detected photons, $\tilde{m}_{+}$, under the projection $\hat{\Pi}{+}$ and the number of detected photons, $\tilde{m}_{-}$, under the projection $\hat{\Pi}_{-}$ can be determined from the peak values of the coincidence counts between SPD 2 and SPD 1, and between SPD 3 and SPD 1, respectively.

In our experiments, we recorded the event timestamps from SPD 1, SPD 2, and SPD 3 over a duration of $\SI{30}{\s}$ for each set of $T$ and $N$. The data were then divided into 600 groups, with each group representing the timestamps of photon counts within $\SI{50}{\ms}$. For each set of $T$ and $N$, we calculated the coincidence counts for these 600 groups between SPD 2 and SPD 1, and between SPD 3 and SPD 1, using a sampling time length of $\SI{50}{\ms}$.

The results are provided in eight individual files, corresponding to an evolution time length $T$ of 2, 4, 6, and 8 for a single-shot evolution ($N=1$), and iteration numbers $N$ of 1, 2, 3, and 4 for a fixed time length ($T=2$). In each file, data labeled as 'Time' represents the delay time axis, data labeled as 'CoinCount1' represents the coincidence counts between SPD 2 and SPD 1 for the 600 groups, and data labeled as 'CoinCount2' represents the coincidence counts between SPD 3 and SPD 1 for the 600 groups.

\begin{figure}[h]
	\centering
	\includegraphics[width=\linewidth]{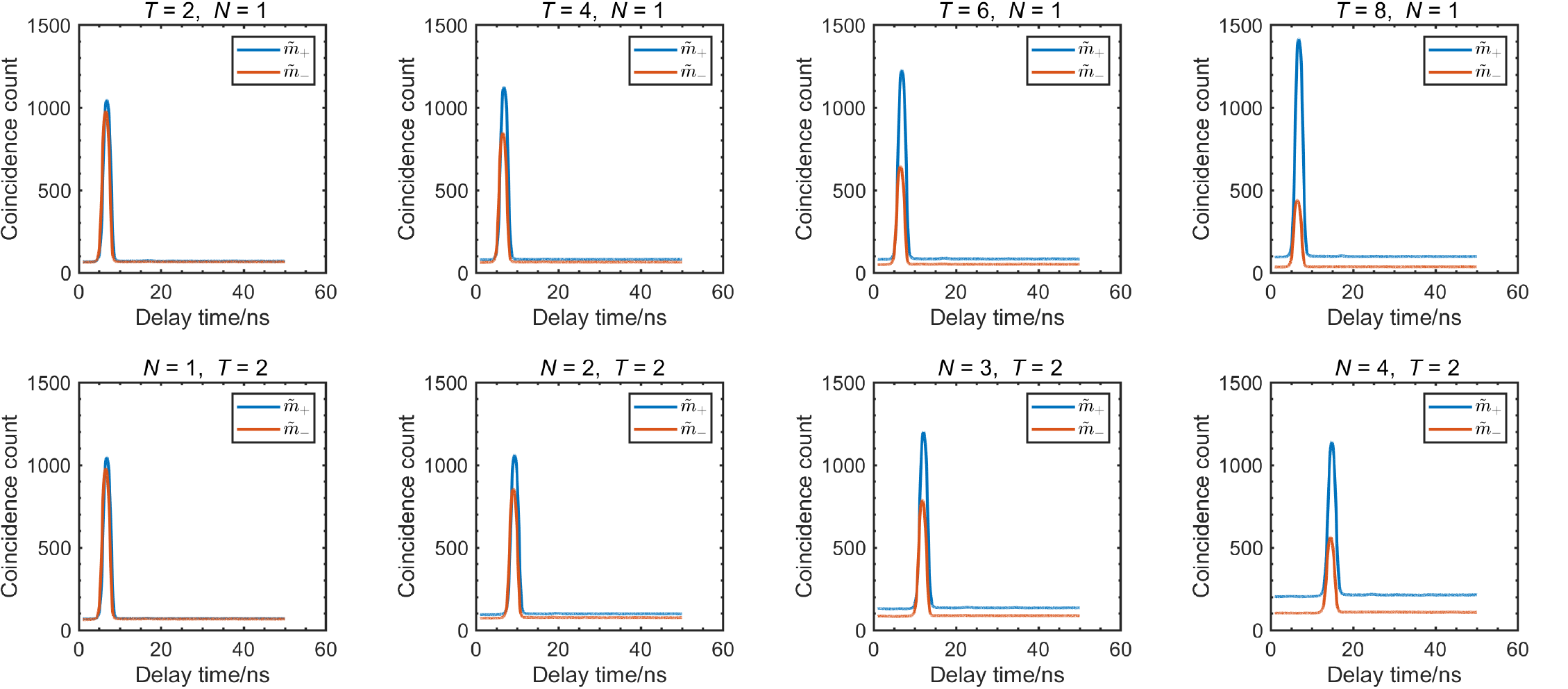}
	\caption{\label{fig:1} Experimental results of coincidence counts with different evolution time length $T$ of the ITD encoding process and the parameterizing process and iteration number $N$ of the identical evolution. The blue curves represent the coincidence counts $\mathcal{C}_{+}$ between SPD 2 and SPD 1, i.e., the detected results under the projection $\hat{\Pi}_{+}$, and the orange curves represent the coincidence counts $\mathcal{C}_{-}$ between SPD 3 and SPD 1, i.e., the detected results under the projection $\hat{\Pi}_{-}$.}
\end{figure}

As shown in Figure~\ref{fig:1}, we present the experimental results of coincidence counts between SPD 2 and SPD 1, and between SPD 3 and SPD 1, for various sets of $T$ and $N$. These results are averaged over the corresponding 600 groups, where the blue curves represent the coincidence counts $\mathcal{C}_{+}$ between SPD 2 and SPD 1, and the orange curves represent the coincidence counts $\mathcal{C}_{-}$ between SPD 3 and SPD 1. The peak values in each curve are marked, which determine the number of detected photons $\tilde{m}_{+}$ under the projection $\hat{\Pi}_{+}$ and the number of detected photons $\tilde{m}_{-}$ under the projection $\hat{\Pi}_{-}$ for each set of $T$ and $N$, respectively.

\bibliography{bibliographyS}